\documentclass[a4paper,twocolumn,accepted=2024-12-17]{quantumarticle}
\pdfoutput=1
\usepackage[utf8]{inputenc}
\usepackage[english]{babel}
\usepackage[T1]{fontenc}
\usepackage{hyperref}
\usepackage{amsmath}

\usepackage{tikz}
\usepackage{lipsum}

\usepackage{graphicx}
\usepackage{dcolumn}
\usepackage{bm}
\usepackage{amsfonts}
\usepackage{braket}
\usepackage{subcaption}
\usepackage{hyperref}

\definecolor{tblue}{rgb}{0.121,0.467,0.706}
\definecolor{orange}{rgb}{1,0.498,0.055}
\definecolor{green}{rgb}{0.172,0.627,0.172}
\definecolor{red}{rgb}{0.839,0.153,0.157}
\definecolor{purple}{rgb}{0.58,0.404,0.741}
\definecolor{brown}{rgb}{0.549,0.337,0.294}

\usepackage{xcolor}
\usepackage{soul}
\renewcommand\st[1]{}
\newcommand{\janos}[2]{\st{#1}{\color{black} {#2}}}
\newcommand{\Aron}[2]{\st{#1}{\color{black} {#2}}}
\newcommand{\Balint}[2]{\st{#1}{\color{black} {#2}}}

\begin{document}

\title{Characterization of errors in a CNOT between surface code patches.}

\author{Bálint Domokos}
 \affiliation{Department of Theoretical Physics, Institute of Physics, Budapest University of Technology and Economics, Műegyetem rkp. 3., H-1111 Budapest, Hungary}
 
\author{Áron Márton}%
\affiliation{Department of Theoretical Physics, Institute of Physics, Budapest University of Technology and Economics, Műegyetem rkp. 3., H-1111 Budapest, Hungary}

\author{János K. Asbóth}
\affiliation{Department of Theoretical Physics, Institute of Physics, Budapest University of Technology and Economics, Műegyetem rkp. 3., H-1111 Budapest, Hungary}
\affiliation{HUN-REN Wigner Research Centre for Physics, H-1525 Budapest, P.O. Box 49., Hungary}%


\begin{abstract}


As current experiments already realize small quantum circuits on error corrected qubits, it is important to fully understand the effect of physical errors on the logical error channels of these fault-tolerant circuits.     
Here, we investigate a lattice-surgery-based CNOT operation between two surface code patches under phenomenological error models. (i) For two-qubit logical Pauli measurements -- the elementary building block of the CNOT -- we optimize the number of stabilizer measurement rounds, usually taken equal to $d$, the size (code distance) of each patch. We find that the optimal number can be greater or smaller than $d$, depending on the rate of physical and readout errors, and the separation between the code patches.
(ii) We fully characterize the two-qubit logical error channel of the lattice-surgery-based CNOT. We find a symmetry of the CNOT protocol, that results in a symmetry 
of the logical error channel. We also find that correlations between X and Z errors on the logical level are suppressed under minimum weight decoding. 
 

\end{abstract}

\maketitle


\section{Introduction}

Fault-tolerant quantum computation will be the basis of future large-scale quantum computers. To protect logical quantum information from environmental noise quantum error correction is essential. Among the array of promising quantum error-correcting codes the surface code \cite{kitaev2003fault,fowler2012surface} stands out as a favored choice due to its high threshold, scalability, and planar connectivity. Recently, surface code memory experiments have been realized on various platforms \cite{google2023suppressing,Bluvstein_2022,Marques_2021,
krinner2022realizing,Zhao_2022}, and the experimental investigation of fault-tolerant quantum gates has started \cite{Bluvstein_2023, zhang2024demonstrating, hetenyi2024creating, menendez2023implementing}.

\begin{figure}[!ht]
    \centering
    \includegraphics[width = .45\textwidth]{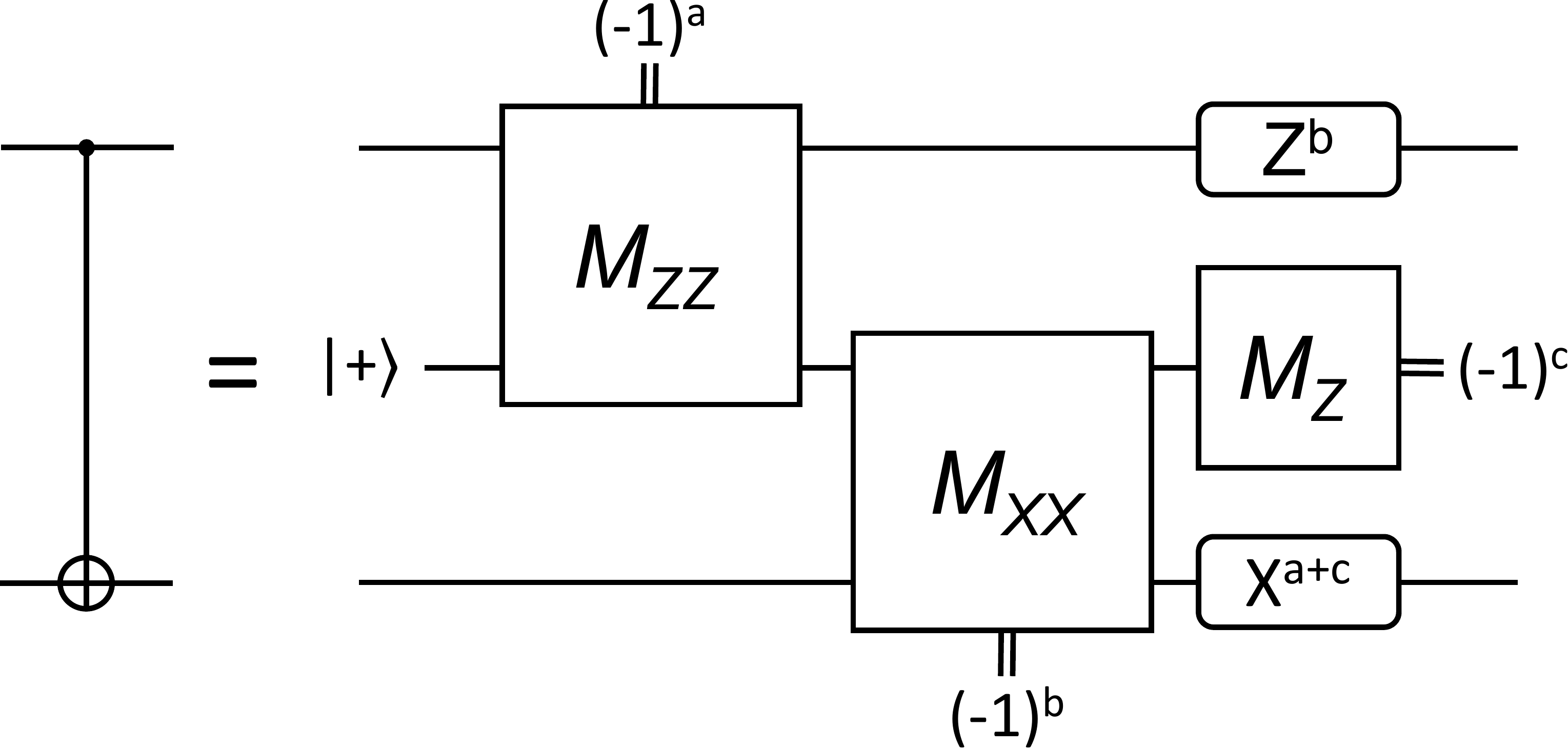}
    \caption{The measurement-based realization of the CNOT operation. This circuit realizes a CNOT gate between the control (top) and target (bottom) qubits, while an ancilla qubit is initialized in the $\ket{+}$ state and at the end of the circuit, measured in the X-basis. Based on the outcomes of mid-circuit measurements Pauli corrections may be performed on the control and target qubits.}
    \label{fig:CNOT_circuit}
\end{figure}

For fault-tolerant universal quantum computation with the surface code a universal set of fault-tolerant gates is required. Single-qubit Clifford gates, such as the Hadamard $H$ and the phase gate $S$, can be implemented by braiding the corners of surface code patches \cite{Brown_2017}. Although transversal multi-qubit logical Clifford operations exist between surface code patches \cite{Shor_1996}, and have been demonstrated recently with neutral atoms \cite{Bluvstein_2023}, this requires high connectivity which is unachievable on several platforms. Another possible way to implement multi-qubit Clifford gates is via measurements of multi-qubit Pauli observables, an example for the CNOT gate \cite{beenakker2004charge} is shown in Fig.~\ref{fig:CNOT_circuit}. Multi-qubit Pauli measurements can be done fault tolerantly with lattice surgery protocols \cite{Horsman_2012,Litinski_2019}, requiring no more than two-dimensional nearest-neighbour connectivity.

Non-Clifford operations can be achieved with Clifford gates and "magic ancillas". Notably, the fault-tolerant preparation of magic states in surface codes involves magic state distillation protocols \cite{Bravyi_2005}, which are quite costly.
Logical magic states with fidelities higher than single qubit magic state fidelities have been distilled with a superconducting architecture \cite{Gupta_2024}. Beyond distillation, numerous other approaches exist to realize universal fault-tolerant quantum computation \cite{Campbell_2017}, including higher-dimensional codes \cite{Brown_2020}, encoding information into islands of qudits \cite{Laubscher_2019}, using gauge color codes instead of surface codes \cite{bombin2015gauge}, and implementing non-Clifford gates through small coherent rotations \cite{choi2023fault}. The costs of these methods have to be compared with state distillation protocols \cite{Beverland_2021}.

Notably, there exist more complex lattice surgery protocols, involving twist defects, or the measurement of more than two logical operators \cite{Litinski_2018,Litinski_2019}, but these schemes are beyond the scope of our work.

In this work, we investigate the lattice-surgery-based fault-tolerant CNOT gate between two surface code patches under phenomenological noise model. This model includes Pauli noise on the physical level and phenomenological readout errors during stabilizer measurements. (i) We show that for finite error rates the optimal number of stabilizer measurement rounds during a ZZ (or XX) measurement can differ slightly from the code distance if the probability of random Pauli X (or Z) errors and the probability of readout errors are the same. We also show that this is not the case for biased physical and readout error rates the optimal number can heavily differ from the code distance. (ii) We fully characterize the logical errors during a CNOT and show that the symmetries of the logical CNOT protocol appear in the two-qubit logical error channel, enforcing the channel to have a special structure. Our findings contribute to a deeper understanding of the noise structure in surface code-based logical operations, and are of interest for experimental implementations of lattice surgery protocols.

The rest of this paper is structured as follows: In Sec.~\ref{sec:surface_code_spacetime_diagrams} we introduce the surface code. We define check generators and the check graph to describe multiple stabilizer measurement rounds. Moreover, we show how to visualize logical protocols with multiple measurement rounds through spacetime diagrams. In Sec.~\ref{sec:ZZ_measurement} we introduce lattice surgery protocols for two-qubit logical Pauli measurements. We optimize the parameters of these  protocols under random Pauli and phenomenological readout noise. In Sec.~\ref{sec:CNOT_characterization} we show that the symmetries of the lattice-surgery-based logical CNOT protocol enforce the logical Pauli channel to have a special structure, 
characterized by 3 logical error parameters. This last result is exact for the independent $X$ and $Z$ errors, and a numerically well-supported approximation for depolarizing noise under minimum weight perfect matching decoding. 

\section{The surface code and spacetime diagrams} \label{sec:surface_code_spacetime_diagrams}



In this paper we investigate the impact of errors on quantum operations between surface code patches -- concepts that we briefly introduce in this Section. \janos{}{We also introduce the three-dimensional mathematical structure used for decoding, i.e., the check graph, and its simplified representation, the spacetime diagrams.} 

\subsection{A single patch of the surface code}
\label{subsec:single_patch}

\begin{figure}[!ht]
    \centering
    \includegraphics[width = .45\textwidth]{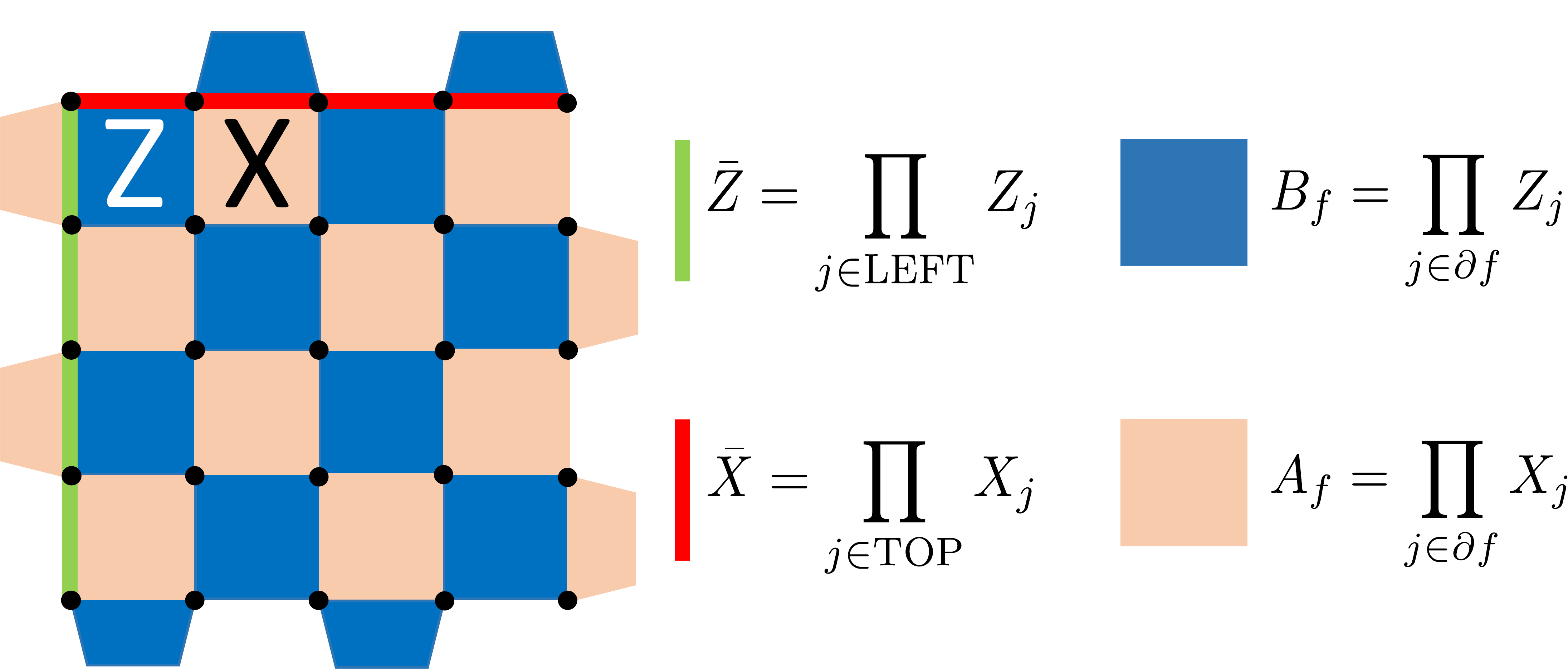}
    \caption{The layout of a distance 5 surface code patch. $Z$- ($X$-) stabilizers are products of $Z$ ($X$) operators acting on the qubits at the corners of dark (light) faces. A logical $Z$ ($X$) operator is a string of single qubit $Z$ ($X$) operators, connecting bottom (left) and top (right) boundaries depicted as a green (red) line.}
    \label{fig:surface_code_layout}
\end{figure}

A (rotated) surface code patch \cite{Bombin_2007} 
is a square grid of $n=d^2$ physical qubits, with $d$ an odd integer, as depicted in Fig.~\ref{fig:surface_code_layout}. 
A single logical qubit is encoded in the collective quantum state of the $n$ physical qubits, in the $+1$ eigensubspace of the $Z$- and $X$- stabilizers (more precisely stabilizer generators), $B_f$ and $A_f$,  
\begin{align}
    B_f &= \prod_{j\in\partial f} Z_j;& 
    A_f &= \prod_{j\in\partial f} X_j,
\end{align}
acting on the qubits at the corners of the corresponding face, $f$. 
The logical states of the code are superpositions of the logical basis states, $\ket{\overline{\psi}} = \alpha \ket{\overline{0}} + \beta \ket{\overline{1}}$, with 
\begin{align}
    \ket{\overline{0}} \!&=\! 
    \prod_{f\in \text{light}} \!\!\! 
    \frac{1+A_f}{\sqrt{2}}\ket{0}^{\otimes n};&
    \ket{\overline{1}} \!&=\! 
    \prod_{f\in \text{light}} \!\!\!
    \frac{1+A_f}{\sqrt{2}}\ket{1}^{\otimes n},
\end{align}
\janos{}{where the subscript "light" denotes faces with light shading, corresponding to $X$-stabilizers.} Thus, $A_f \ket{\overline{\psi}} = B_f \ket{\overline{\psi}} = \ket{\overline{\psi}}$ for all logical states $\ket{\overline{\psi}}$.

On the left and right (vertical) boundaries of the surface code patch, there are only $X$-stabilizers, while on the top and bottom (horizontal) boundaries, only $Z$ ones. Therefore, we call these $X$- and  $Z$-boundaries; these are also known as smooth and rough boundaries.

Logical Pauli operators $\overline{X}$ and $\overline{Z}$ are realized by strings of $X$ and $Z$ operators connecting different $X$- and $Z$-boundaries -- examples are in Fig.~\ref{fig:surface_code_layout}. They commute with the stabilizers, moreover, these operators multiplied by stabilizers are also logical operators, acting equivalently in the logical subspace. 
The code distance $d$ of a patch is the length (weight) of the shortest logical operator, equal to the linear size of the patch, i.e., $d=\sqrt{n}$.

\subsection{Errors, stabilizer measurements, check graphs}

We consider two types of errors, against which the surface code protects the encoded qubit \Aron{}{during a logical operation}: data qubit errors and phenomenological readout errors. We restrict data qubit errors to probabilistic single-qubit Pauli $X$ and $Z$ (if these occur on the same qubit, they result in Pauli $Y$). Phenomenological readout errors are perfectly performed measurements with erroneously reported results. 
Both types of errors can be detected by repeated measurements of the stabilizer operators. 
We will refer to the errors as
\begin{align*}
\begin{split}
    &\text{\textbf{Spacelike error}: Pauli error on a data qubit;} \\
    &\text{\textbf{Timelike error}: Readout error on a stabilizer.}
\end{split}
\end{align*}

\begin{figure}[!h]
    \centering
    \includegraphics[width=0.5\linewidth]{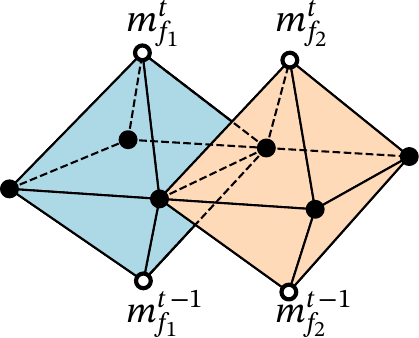}
    \caption{\Balint{}{$X$-type and $Z$-type check generators consist of the stabilizer measurements at time $t$ and $t-1$. The value is determined by the sum of the outcomes, as per Eq.~\eqref{eq:check_gen},  
    and can detect errors on the physical qubits and virtual bits.}}
    \label{fig:check_gens}
\end{figure}

\janos{}{To detect errors, we need to repeatedly measure the $Z$- and $X$- stabilizers; this is represented by \emph{virtual bits}. 
In Figs.~\ref{fig:check_gens}, \ref{fig:time_boundaries}, the virtual bit obtained from a measurement at time $t$ is  a white dot after the layer at time $t$ of physical qubits (black dots).}
Spacelike errors are located on physical qubits and timelike errors are on virtual \janos{}{bits (an erroneous stabilizer measurement is a bit-flip on a virtual bit)}.


The changes that errors cause in the measured Z- and X- stabilizer values are represented by \emph{check generators} (also called \emph{detectors} \cite{McEwen_2023,Gidney_2021_stim}). 
These are generalizations of the concept of stabilizers
: collections of measurement outcomes that should be compared to signal locations of errors. For intermediate time steps during logical operations, a check generator consists of consecutive stabilizer measurement outcomes, as in Fig.\ref{fig:check_gens}, and its value is 
\begin{align} \label{eq:check_gen}
    s_f^t = m_f^t + m_f^{t-1} \quad (\text{mod 2}), 
\end{align}
where $m_f^t$ is the outcome of the measurement of $A_f$ (or $B_f$) at time $t$. 
If data qubit measurements or initializations at time $t$ or $t-1$ are involved, the check generators slightly differs from Eq.\eqref{eq:check_gen}, as we will discuss later.

\Aron{}{All the information from the check generators obtained during a logical operation, performed using the surface code, is represented by a \emph{check graph.}} \Aron{}{This graph} has the check generators at the nodes and the possible errors (physical qubits and virtual bits) as edges. \Aron{}{Each edge has a weight
\begin{align}
    w_e = \ln\Big(\dfrac{1-p_e}{p_e}\Big),
\end{align}
where $p_e$ is the probability of a spacelike/timelike error on the given data qubit/virtual bit.} 

The check graph consists of two disconnected pieces: the $X$-check graph contains $X$-check generators (comparing measurement outcomes of $X$-stabilizers), and the $Z$-check graph contains $Z$-check generators. $X$-type errors can violate $Z$-check generators, while $Z$-type errors can violate $X$ ones. Notably, correlations between $X$- and $Z$- spacelike errors can occur due to Pauli $Y$ noise. However, no such correlations occur on timelike errors, because readout errors on $Z$- and $X$- stabilizers ($X$- and $Z$- type timelike errors) are independent.

To infer errors from the measurement data in the check graph (the syndrome) a decoding algorithm has to be run on \Aron{the check graph}{it}; we took the minimum-weight perfect matching (MWPM) \cite{dennis2002topological,fowler2014minimum} approach. 
This produces a correction string for any given syndrome, which has both timelike and spacelike corrections. We used the efficient Blossom algorithm \cite{edmonds_1965} to find the minimum-weight string, as implemented in the PyMatching software package \cite{higgott2022pymatching,higgott2023sparse}.
This solves the decoding on $X$- and $Z$-check graphs independently, with proper weighting of the edges to account for different data qubit and readout error rates. Our codes and the numerical data are available at \cite{data}

\subsection{Timelike and spacelike boundaries of check graphs}

\begin{figure}[!ht]
    \centering
    \includegraphics[width = .45\textwidth]{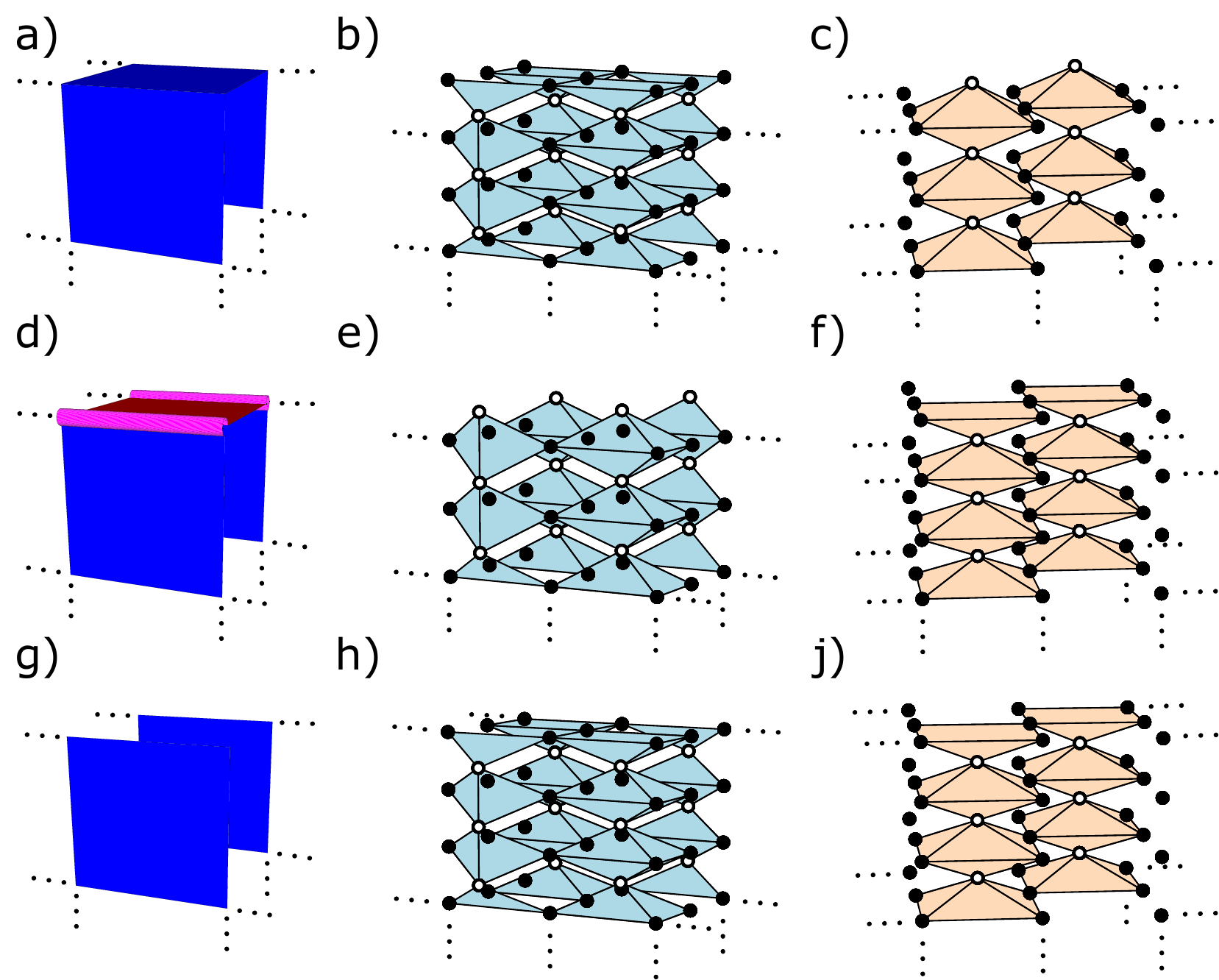}
    \caption{Spacetime diagram segments and the corresponding check graphs representing the three different timelike boundaries. Black (white) dots: physical (virtual) qubits (bits), blue (orange) volumes: $Z$- ($X$-) check generators. Top row: upper timelike $Z$-boundary, with a) spacetime diagram, b) $Z$-check graph, c) $X$-check graph. Here only $Z$-check generators are present at the top boundary. \Balint{}{There is an extra layer of physical qubits on the $Z$-check graph due to the data qubit readout.} Middle row: upper timelike $X$-boundary, with d) spacetime diagram, e) $Z$-check graph, f) $X$-check graph. Here only $X$-check generators are present at the top boundary. Bottom row: upper perfect timelike boundary, with g) spacetime diagram, h) $Z$-check graph, j) $X$-check graph. These both contain check generators at the top boundary.}
    
    
    \label{fig:time_boundaries}
\end{figure}

\janos{}{We must pay special attention to (2-dimensional) boundaries of (3-dimensional) check graphs, both spacelike and timelike boundaries, both of which can be $X$-type or $Z$-type.
These are generalizations of the (1-dimensional) $X$- and $Z$-boundaries of a single (2-dimensional) patch of surface code  in Sect.~\ref{subsec:single_patch}. }
Examples are shown in Fig.~\ref{fig:time_boundaries}. 

\emph{Spacelike boundaries} can be defined as surfaces containing the same boundary types in each stabilizer measurement round. Differently put, spacelike $X$- ($Z$-) boundaries are vertical surfaces, where only $X$- ($Z$-) check generators are present. 


\emph{Timelike lower boundaries}, of X- or Z-type, occur in a check graph on areas where physical qubits are initialized, in $\ket{0}$, or in $\ket{+}$, respectively.  
In this case, the values of $Z$- (respectively, $X$-) stabilizers are predetermined before the first round of stabilizer measurements. Therefore, $Z$- (respectively, $X$-) check generators at such a boundary contain \Aron{}{only single stabilizer measurements, and their values are given as:
\begin{align}
    s_f^{t_0} = m_f^{t_0},
\end{align}
where the qubits were initialized directly before time $t_0$. 
}\Aron{physical qubits from the "zeroth" time step and virtual qubits from the first stabilizer measurement round.}{} In contrast, the measurement outcomes of $X$- (respectively, $Z$-) stabilizers are random, therefore there is a missing layer of $X$- (respectively, $Z$-) check generators at this boundary. 

\emph{Timelike upper boundaries} correspond to  measurements of data qubits in some region in the $Z$- ( respetively, $X$-) basis at some time $t_f$. \Aron{}{In that case the values of $Z$- (respectively, $X$-) stabilizers can be constructed from the data qubit measurements. Therefore, $Z$- (respectively, $X$-) check generators at such boundaries contain the stabilizer measurements from the $t_f - 1$-th step, and also the corresponding data qubit measurements from the $t_f$-th step. The values of such check generators are given as:
\begin{align}
    s_f^{t_f} = m_f^{t_f-1} + \sum_{j\in \partial f}m_j \quad (\text{mod 2}),
\end{align}
where $m_j$-s are the data qubit measurement outcomes.}
Thus, similarly to lower timelike boundaries, these boundaries have an extra layer of $Z$-check generators with resepect to $X$-generators (respectively, extra layer of $X$- check generators).

\emph{Perfect timelike boundaries} are timelike boundaries of check graphs with both check generator types present. This scenario cannot be constructed by any specific data qubit initialization/measurement pattern. However, this is a useful artificial construction for the investigation of fault-tolerant logical protocols. 

We illustrate the three different timelike boundaries \Aron{}{with the corresponding check graphs} in Fig.~\ref{fig:time_boundaries}. 

Error strings can terminate at boundaries (both spacelike and timelike) without violating check generators. $X$-type strings can terminate at $X$ boundaries while $Z$-type strings at $Z$ boundaries. We define logical errors through the boundaries as well: an error string is a logical error string (can consist of both spacelike and timelike errors) if it connects two disconnected boundaries, from the same type, without violating any check generators. We also define the so-called \emph{fault distance} of a logical protocol, which is the length of the shortest logical error string. This is a straightforward generalization of code distance for logical protocols. 

\subsection{Spacetime diagrams: visualization of protocols}

\janos{}{For a simple overview of a realization of a logical operation using the surface code, one can represent only the boundaries of the check graph in a so-called \emph{spacetime diagram}}
\cite{Bombin_2023}, as in Fig.~\ref{fig:time_boundaries}. 
We will use this simplified representation often below, as  
spacetime diagrams are useful tools to identify different logical error strings and follow their propagation through the protocol.

We note that spacetime diagrams which represent general logical protocols can have a richer repertoire of topological features \cite{gidney2023inplace,Bombin_2023,gehér2023errorcorrected}. If so-called domain walls are present twists can have other functions than just separating different boundary types, and logical error strings are somewhat harder to define. Generally, spacetime diagrams can represent all surface code-based logical Clifford operations \cite{Bombin_2023}.

\section{Lattice-surgery-based ZZ-measurements: optimization of the number of measurement rounds}
\label{sec:ZZ_measurement}

Lattice surgery \cite{Horsman_2012,Chamberland_2022_1,Chamberland_2022_2} is one of the leading approaches to fault-tolerant quantum computation using surface code patches. 
The basic two-qubit operation of lattice surgery is 
%
the measurement of the logical $\Bar{Z}\Bar{Z}$ or the $\Bar{X}\Bar{X}$ operator. 
In this Section we 
review how the measurement of the $\Bar{Z}\Bar{Z}$ operator is performed, and investigate numerically how to optimize the number of measurement rounds for this two-qubit logical operation. 
The results for the $\Bar{X}\Bar{X}$ measurement can be obtained by straightforward generalization. 

We consider two surface code patches of distance $d$ with $X$-boundaries facing each other, separated by a "bridge" of $w$ columns of data qubits (coupling qubits). In practice, $w$ can vary, depending on the physical distance of the two patches on the quantum computer; with some patch definitions $w=0$ is also feasible \cite{Litinski_2018}. 

The steps of the $\Bar Z \Bar Z$-measurement protocol are:
1) Initialize all coupling qubits in $\ket{+}$;
2) Merge the patches, by measuring the stabilizers of the extended rectangular patch; 3) Measure the coupling qubits in the $X$-basis. 

Without any errors, the outcome of the $\Bar{Z}\Bar{Z}$ measurement is the product of the measurement outcomes of the newly defined intermediate $Z$-stabilizers. 

Measurement errors on the intermediate stabilizers might lead to errors in the result of the $\Bar{Z}\Bar{Z}$ measurement. Thus, for fault tolerance, multiple stabilizer measurement rounds are needed. The criterion for ensuring a fault-tolerant measurement with a fault distance $d$ is to perform at least 
\begin{equation}
    h_2 = d
\end{equation}
rounds of stabilizer measurements between merge and
split operations.



\begin{figure}[!ht]
    \centering
    \includegraphics[width = .6\columnwidth]{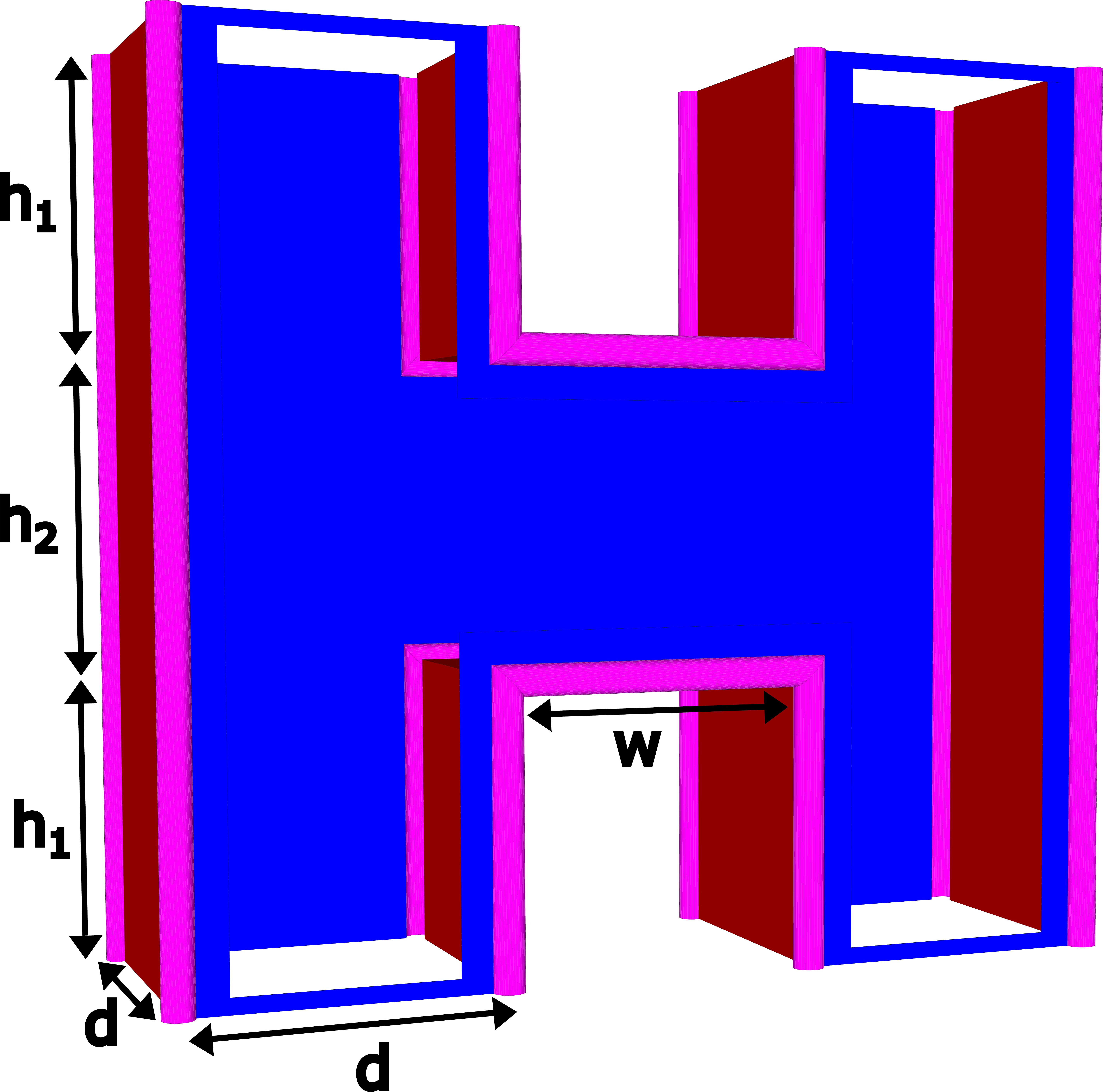}
    \caption{The spacetime diagram of the lattice-surgery-based $\Bar{Z}\Bar{Z}$ measurement. Parameters are the linear size $d$ (in data qubits) of each patch, and the length $w$ of the "bridge" between them; and the number of measurement rounds $h_1$ before/after, and  $h_2$ during the phase where the patches are merged.}
    \label{fig:ZZ_spacetime_diagram}
\end{figure}

The spacetime diagram of the $\Bar{Z}\Bar{Z}$ measurement is shown in Fig.~\ref{fig:ZZ_spacetime_diagram}. This diagram has perfect timelike boundaries at the top and at the bottom, and timelike $X$-boundaries in the middle. It consists of 4 disconnected $X$- and 2 disconnected $Z$-boundaries.

\subsection{Optimization of the number of measurement rounds in a two-qubit Pauli measurement}

\begin{figure}[!ht]
    \centering
    \includegraphics[width=.48\textwidth]{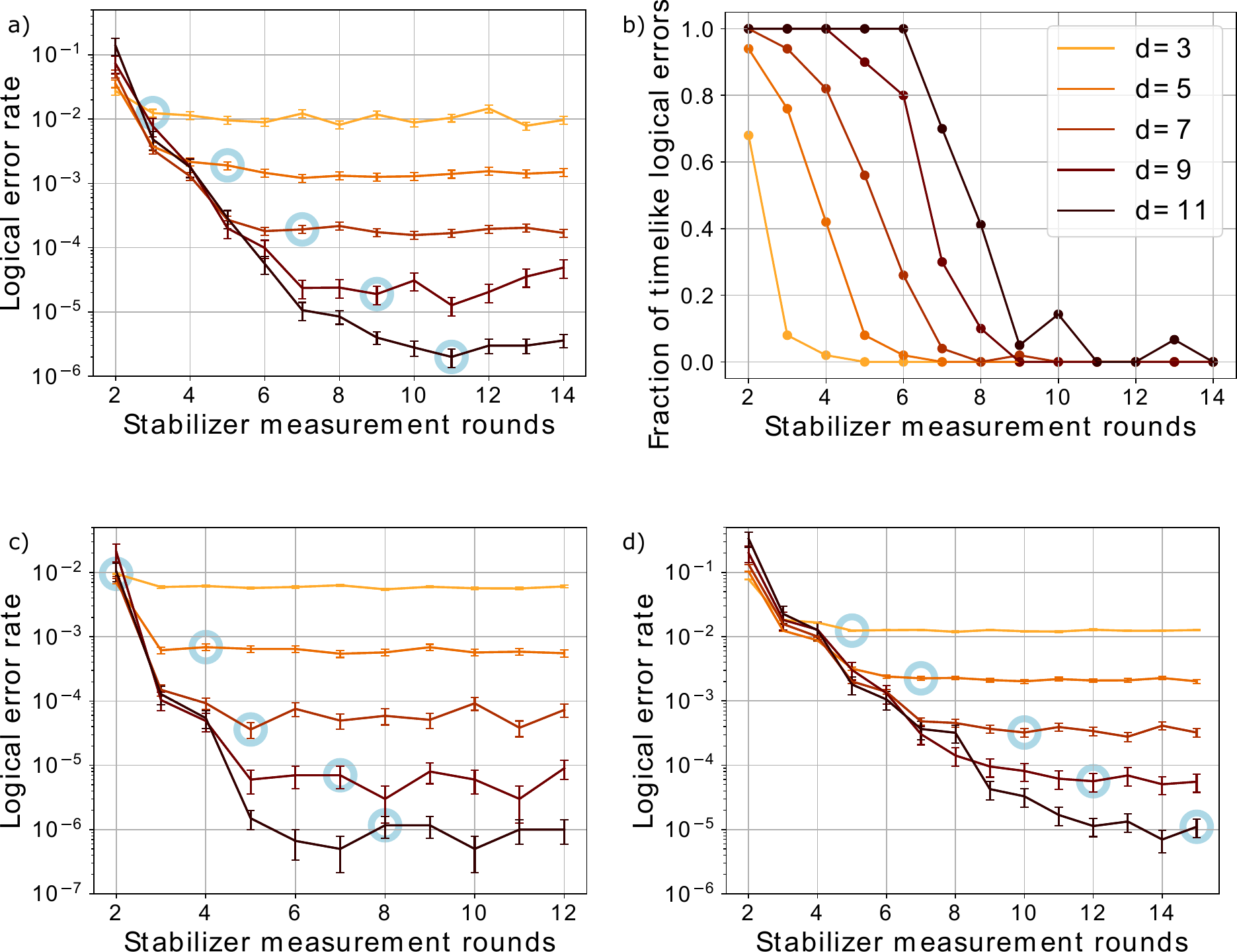}
    \caption{Errors of the $\Bar{Z}\Bar{Z}$ measurement as a function of $h_2$, the measurement rounds in the merged phase, with $d=7$, $h_1 = h_3 = 1$, $w = 1$, and data qubit error rate $p=0.6\%$. We find that  Eq.\eqref{eq:optimal_h2} (encircled values) is a decent, but not optimal choice, for readout error (a) $q=p$,  (c) $q=p/6$, and (d) $q=3p$. (b): The fraction of timelike logical errors for different $h_2$ values for  $q=p$. }
    \label{fig:error_decays}
\end{figure}

What is the optimal number of measurement rounds of the stabilizers in the phase when the two patches are merged, $h_2$?  
Increasing $h_2$ suppresses the contribution of timelike errors to the probability of a logical error, at the cost of an overhead: longer logical gate operation, and a slight increase in the probability of spacelike errors leading to a logical error. 
If $h_2$ is quite long, the contribution of timelike errors to the logical error is already tiny, and thus, no big gains are expected from increasing $h_2$ (the increased probability of spacelike errors can make the logical error rate even slightly worse). Thus, the optimal number of rounds will be the value of $h_2$ where the logical error rate saturates (no more practical advantage in increasing $h_2$).
This optimal number of rounds is often taken to be $h_2 \approx d$, but this can depend on the geometrical parameters and the error rates as well. We aim to find how this depends on, e.g., the patch size $d$, the length of the bridge $w$, 
and on the error rates. 


We investigated the optimal value of 
$h_2$, with 
qubit noise that is maximally biased in $X$, with spacelike error probability $p$: 
\begin{equation} \label{eq:simplifiederrorchannel}
    \varepsilon(\rho) = (1-p)\rho + p(X\rho X), 
\end{equation}
and timelike error probability $q$.

We first estimate the optimal value of $h_2$ analytically, by a series expansion treating the error rates as small parameters. 
Taking into account only the lowest order terms in spacelike and timelike error rates $p$ and $q$, the logical error rate is
\begin{equation} \label{eq:P_L_p_q}
    P_L(p,q) = Ap^{(d+1)/2} + Bq^{(h_2+1)/2} + \ldots .
\end{equation}
Here $A$ ($B$) is the multiplicity of the shortest spacelike (timelike) error strings that can lead
to a logical error. 
Higher order terms are strongly suppressed in the case of small error rates.
We see that $h_2$ reaches its optimal value when the competing terms in Eq.~\eqref{eq:P_L_p_q} are equal. From that on it is not that beneficial to further increase $h_2$, because the spacelike error mechanisms are already the dominant sources. 

Thus, the optimal value for $h_2$ in the small $p,q$ limit is
\begin{equation}
\label{eq:optimal_h2}
    h_2 \approx (d+1)\dfrac{\ln (p)}{\ln (q)} -1,
\end{equation}
where we omitted the term 
\begin{equation} \label{eq:ln_term}
    \frac{2\text{ln}(A/B)}{\text{ln}(q)},
\end{equation}    
assuming that in the $q \rightarrow 0$ limit it is negligible.
We note that this is analogous to arguments showing that elongated rather than square shaped surface code patches are advantageous for biased noise \cite{Bonilla_Ataides_2021}.

\begin{figure}[ht]
    \centering
    \includegraphics[width = .45\textwidth]{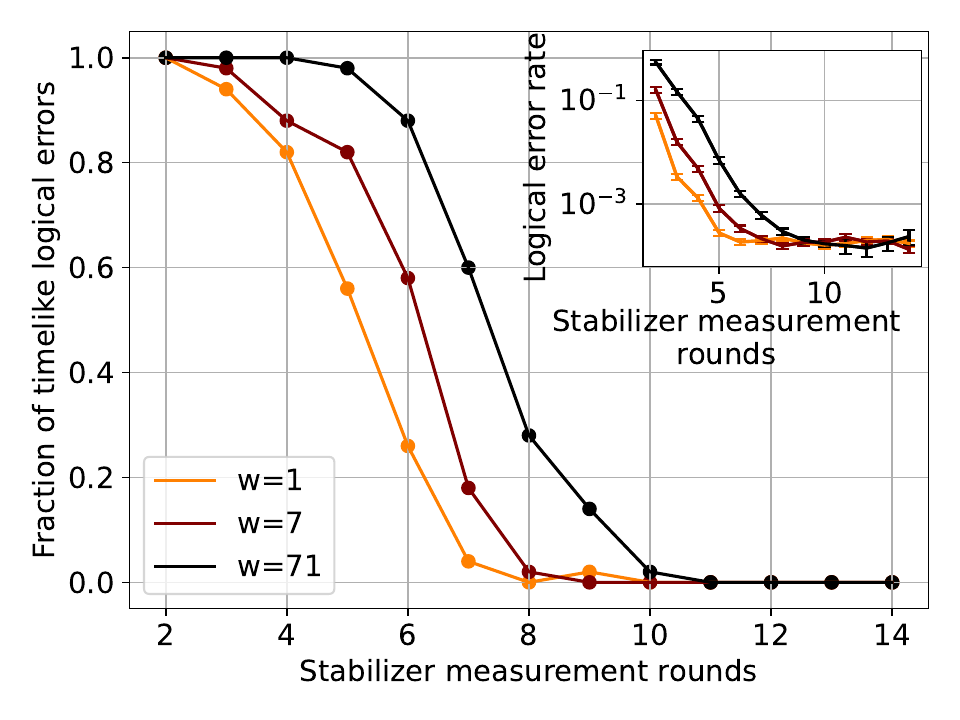}
    \caption{The fraction of timelike logical errors as a function of the number of measurement rounds during the merged phase($h_2$), for different bridge lengths $w$. Here, $d=7$, and $p=q=0.006$. The is shifted from $h_2$=7 to smaller values for $w=1$, higher values for large $w$, due to the higher multiplicity of timelike errors.}
    \label{fig:w_increase}
\end{figure}

We sampled logical errors via Monte Carlo simulations, by running minimum weight perfect matching decoder (PyMatching) \cite{higgott2022pymatching,higgott2023sparse} on the $X$- and $Z$-check graphs of the $\Bar{Z}\Bar{Z}$ measurement. Our results show that the analytical approximation of Eq.~\eqref{eq:optimal_h2} slightly overestimates the optimal value of $h_2$. 
We simulated the $\Bar{Z}\Bar{Z}$ measurement with geometrical parameters $d=7, w=1, h_1 = h_3 = 1$ for three cases: $p = q$, $p = 6q$, $3p = q$.
As shown in Fig.~\ref{fig:error_decays}, in all three cases, we see that the decrease of the logical error rate slows down significantly for smaller $h_2$ values than Eq.\ref{eq:optimal_h2} suggests. 
For a more detailed picture, we also show the ratio of timelike to logical errors, which transitions from 1 to 0 as $h_2$ is increased. The critical value of $h_2$ where the transition happens is smaller than expected based on the $p,q\to 0$ limit. This suggests that the 
correction in Eq.\eqref{eq:ln_term} remains significant with these geometrical ($w=h_1=1, d=7$) and error parameters ($p,q\approx 10^{-3}$).

We found that increasing the separation $w$ of the patches the optimal value of $h_2$ for the $\Bar{Z}\Bar{Z}$ measurement increases slightly, as shown in Fig.~\ref{fig:w_increase}. 
This is due to the increasing multiplicity of the shortest timelike logical error string, while the multiplicity of the spacelike logical error string does not change. The optimal value of $h_2$ is where the fraction of timelike logical errors drops. Note that the increase in the optimal value of $h_2$ is quite small even for distant patches ($w\approx10d$): previous estimates predicted a logarithmic growth \cite{Litinski_2017}.

In case $Z$ errors are included in the error model we expect qualitatively similar results, but the optimum should come at an even lower $h_2$ than in Fig.~\ref{fig:error_decays}. 
Here the multiplicity of spacelike logical $Z$ error strings will be much higher than that of other spacelike error strings.
Moreover, the multiplicity of spacelike and timelike logical error stings increase together in the case of an increasing $w$, hence we expect no shift of the transition in the fraction of timelike logical errors. 

\section{Investigation of lattice-surgery-based CNOT protocol}
\label{sec:CNOT_characterization}

\begin{figure}[ht]
    \centering
    \includegraphics[width = .6\columnwidth]{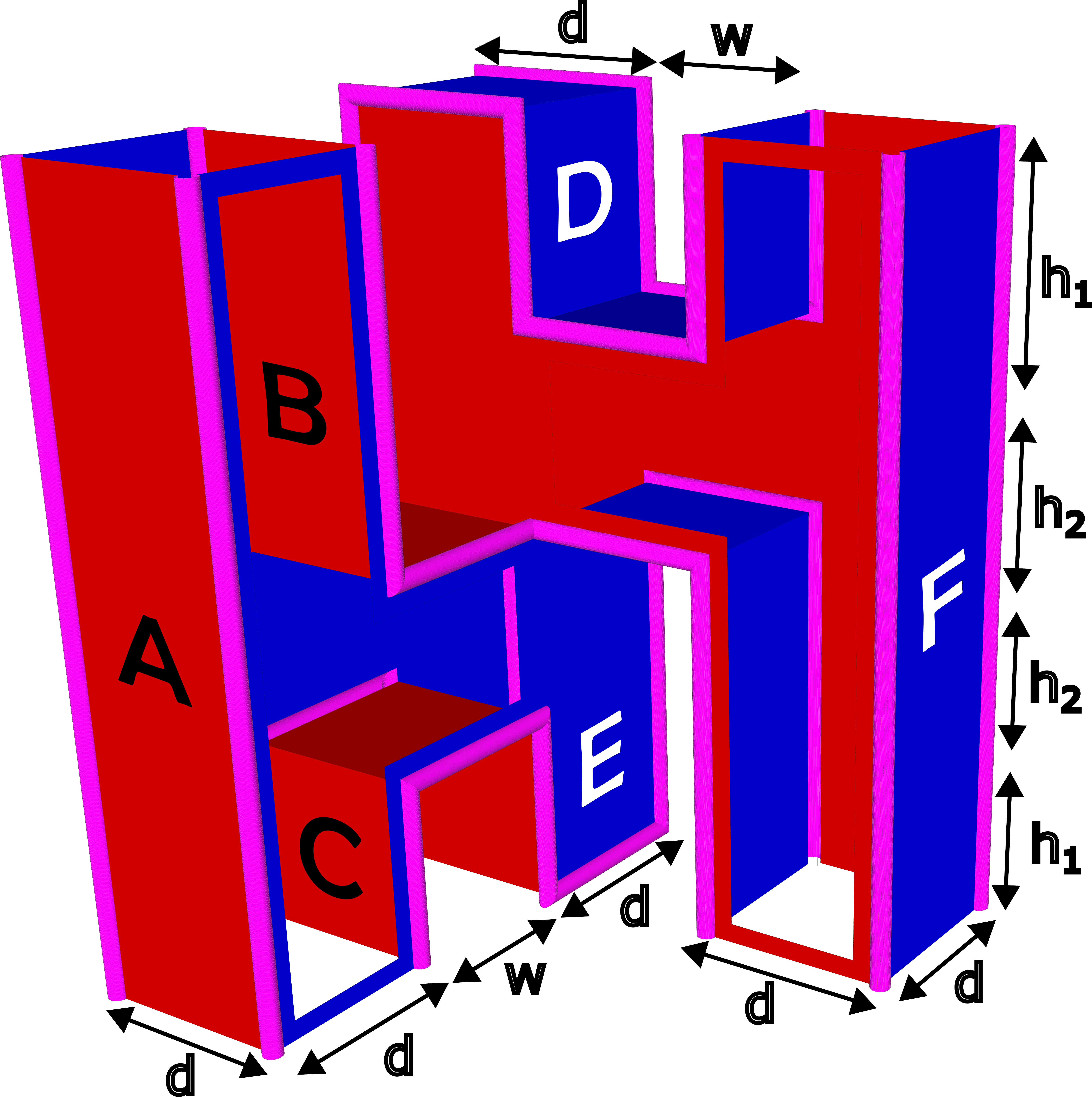}
    \caption{Spacetime diagram of the lattice-surgery-based CNOT. Control (left) and target (right) patches have perfect lower and upper timelike boundaries, ancilla patch (middle) has lower timelike $X$- and upper timelike $Z$-boundaries: initialized in $\ket{+}$, measured in $Z$. The $X$- ($Z$-) boundaries consist of three disconnected parts labeled $A,B,C$ ($D,E,F$). 
    }
    \label{fig:CNOT_spactime_diagram}
\end{figure}

In this Section we fully characterize the two-qubit Pauli logical error channel of the standard lattice-surgery-based CNOT protocol \cite{Horsman_2012}, assuming single qubit Pauli noise on the data qubits and phenomenological readout errors. This characterization amounts to analytical and numerical estimates for the probabilities of the 15 possible two-qubit logical Pauli errors that can occur during a CNOT.  
Note that this CNOT gate follows the logical circuit of Fig.~\ref{fig:CNOT_circuit}. 

The spacetime diagram of this CNOT, 
Fig.~\ref{fig:CNOT_spactime_diagram}, 
has three disconnected $X$-boundaries, labeled by $A$, $B$, $C$, and three disconnected $Z$- boundaries: $E$, $F$, $G$.

Any string operator connecting two disconnected boundaries of the same kind leads to a Pauli error on the logical level. 
Since any of these errors, squared, gives the identity,  
there are $16$ distinct connection classes based on the parity of string endings at boundaries $A$, $B$, $C$, $D$, $E$, $F$. Note that a string connecting $BC$ is equivalent to a pair of strings connecting $AB$ and $BC$, and likewise for $EF$, $FG$, and $EG$. These connection classes correspond to the 16 inequivalent two-qubit logical Paulis (includng the error-free case) incurred during the CNOT. The correspondence between string ending parities, connection classes and logical Paulis is summarized in Table~\ref{tab:Pauli_strings_classes}. 
\begin{table}[!ht]
\centering
\begin{tabular}{|ccc|c|c|}
    \hline
    $N_A$ & $N_B$ & $N_C$ & Connection & Logical Pauli  \\
    \hline
    0 & 0 & 0 & $\emptyset_{ABC}$ & 1 \\
    1 & 1 & 0 & A - B & $X_1$ \\
    1 & 0 & 1 & A - C & $X_1X_2$ \\
    0 & 1 & 1 & B - C & $X_2$ \\
    \hline
    $N_D$ & $N_E$ & $N_F$ & Connection & Logical Pauli \\
    \hline
    0 & 0 & 0 & $\emptyset_{DEF}$ & 1 \\
    1 & 1 & 0 & D - E & $Z_1$ \\
    1 & 0 & 1 & D - F & $Z_2$ \\
    0 & 1 & 1 & E - F & $Z_1Z_2$ \\
    \hline
    \end{tabular}
    \caption{Correspondence between logical Pauli errors and connection classes of the boundaries. First column: the number of error string endings modulo 2 at each  boundary specifies the connection class -- the sum of every row must be even, since each string has two ends. Second column: The minimally required connections between boundaries in each class, $\emptyset_{ABC}$ ($\emptyset_{DEF}$) signals no connections between $X$- ($Z$-) boundaries. Third column: the corresponding logical Paulis. The 16 logical Pauli error classes can be obtained by multiplying 1 row from the top and 1 from the bottom half of the table.}
    \label{tab:Pauli_strings_classes}
\end{table}

We examine two error models: independent $X$ and $Z$ errors and depolarizing noise, in both cases combined with phenomenological readout errors with rate $q$. 
In the first case, we set the three error rates equal:
\begin{align}
    \label{eq:indep_channel}
    \varepsilon(\rho) &= (1-p)^2\rho + p(1-p)(X\rho X+ Z\rho Z) + p^2Y\rho Y
    ; \nonumber \\  
    q &= p.
\end{align}
In the second case we set $q$ to a value that ensures equal timelike and spacelike error rates on both the $X$ and $Z$ check graphs, 
\begin{align} \label{eq:depol_channel}
    \varepsilon(\rho) &= (1-p)\rho + \dfrac{p}{3}(X\rho X + Z\rho Z  + Y\rho Y); \nonumber \\
    q &=\dfrac{2}{3}p.
\end{align}

\subsection{Symmetry of the CNOT protocol and its consequences on the logical errors}
\label{subsec:Symmetry}


\begin{figure}[!ht]
    \centering
    \includegraphics[width = .45\textwidth]{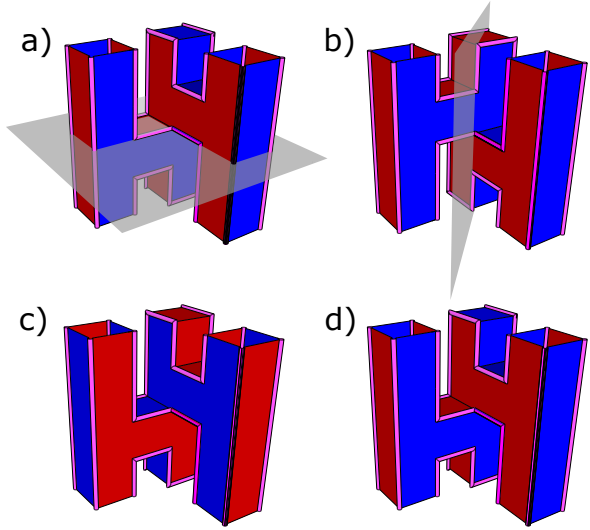}
    \caption{Symmetry transformation of the 
    spacetime diagram of the CNOT (a), in 3 steps. 
    First, time reversal, resulting in (b); Then, mirror reflection, resulting in (c); finally, swapping the $X$ and $Z$ labels (colors), resulting in (d), identical to (a).}
    \label{fig:CNOT_CPT}
\end{figure}

\begin{figure}[!ht]
    \centering
    \includegraphics[width=.48\textwidth]{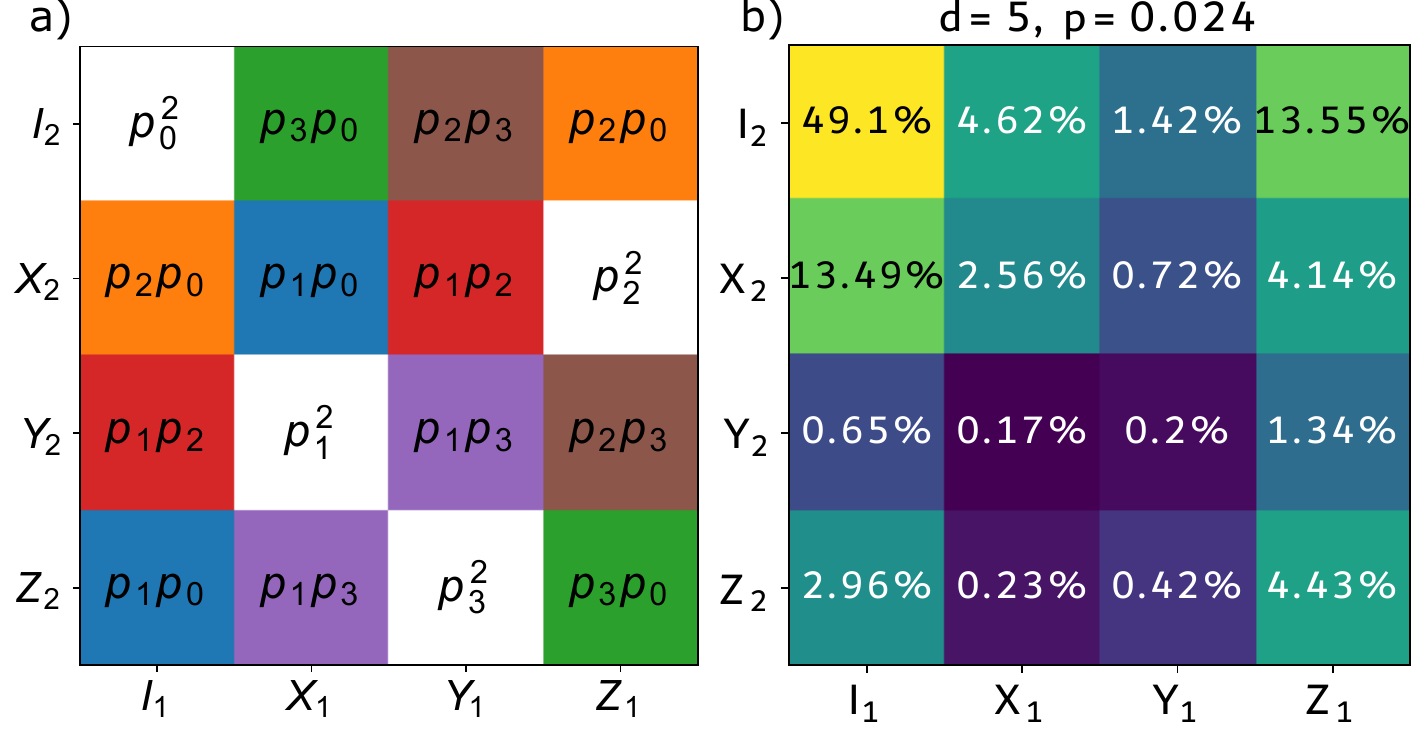}
    \caption{The structure of the two-qubit logical error channel of the lattice-surgery-based CNOT for independent noise. a) Theoretical expectation for the two-qubit logical Pauli error probabilities based on the connection parameters $p_1, p_2, p_3$; non-white colors indicate symmetry partners. b) A numerically obtained two-qubit Pauli error channel that shows  the expected symmetry pattern; colors indicate error rate as also written in the boxes.}
    \label{fig:table_heatmap}
\end{figure}

The CNOT protocol has a symmetry, evident from the spacetime diagram in Fig.~\ref{fig:CNOT_CPT}, which constrains the possible logical error rates. The symmetry 
is composed of: 1) a time reversal, 2) a mirror reflection on a vertical symmetry plane of the "L" shape, and 3) flipping the $X$ and $Z$ labels. As a result the original check graph effectively does not change. The order of the three steps is interchangeable.
If this symmetry transformation also leaves the spatial distribution of error probabilities invariant (as in the case of independent and depolarizing noise), the logical Pauli error strings which transform into each other under the symmetry have to have the same probabilities. As a consequence, symmetry partners among logical Pauli errors have the same probabilities, resulting in a two-qubit logical error channel shown in Fig.~\ref{fig:table_heatmap}a).



Due to the symmetry of the CNOT protocol, in case of independent $X$ and $Z$ errors with equal rate, the probabilities of the different error classes factorize. Therefore, all error class probabilities can be written using the probabilities of three possible classes of connections of different boundaries of one kind via error + correction strings:
\begin{subequations}
\begin{align}
    p_1 &= \mathbb{P}(A - C) = \mathbb{P}(E - F)  ;\\
    p_2 &= \mathbb{P}(B - C)  = \mathbb{P}(D - E) ;\\
    p_3 &= \mathbb{P}(A - B) = \mathbb{P}(D - F) ,
\end{align}
\end{subequations}
using the notation of Table~\ref{tab:Pauli_strings_classes}. 
The probability that all $Z$-boundaries have an even number of $Z$-string endings (equal to the corresponding probability for all $X$-boundaries) can also be written using $p_1$, $p_2$, and $p_3$, as
\begin{equation}
    p_0 = \mathbb{P}(\emptyset_{ABC}) = \mathbb{P}(\emptyset_{DEF}) = 1-p_1-p_2-p_3.
\end{equation}
The table of probabilities of logical Pauli errors building on this connection is shown in Fig.~\ref{fig:table_heatmap}a). 



\subsection{Numerical characterization of the logical error channel}


\begin{figure}[!ht]
    \centering
    \includegraphics[width = .9\columnwidth]{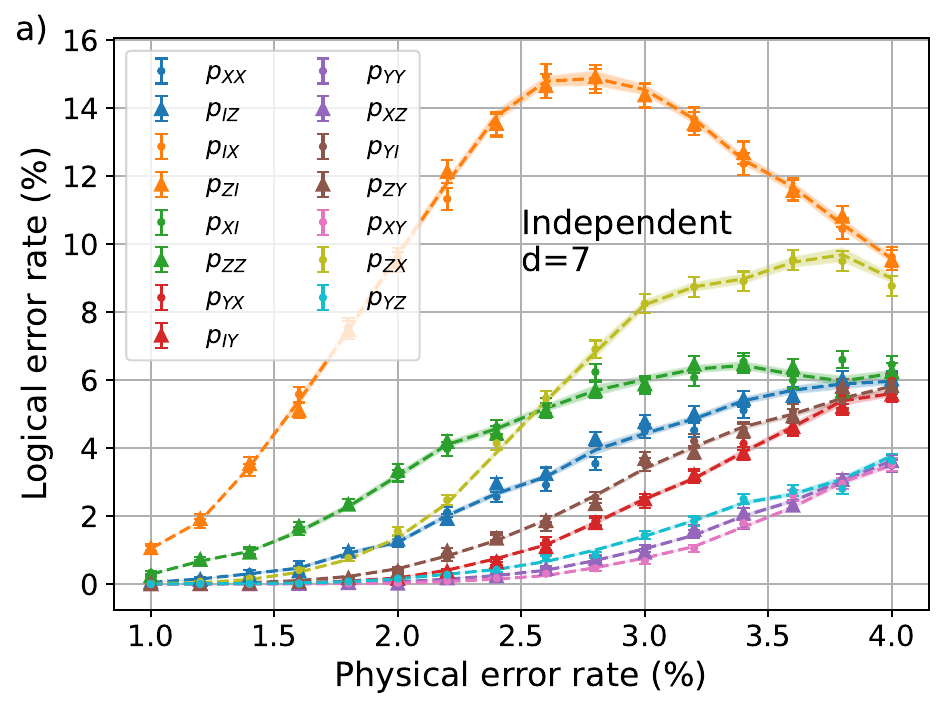}
    \includegraphics[width = .9\columnwidth]{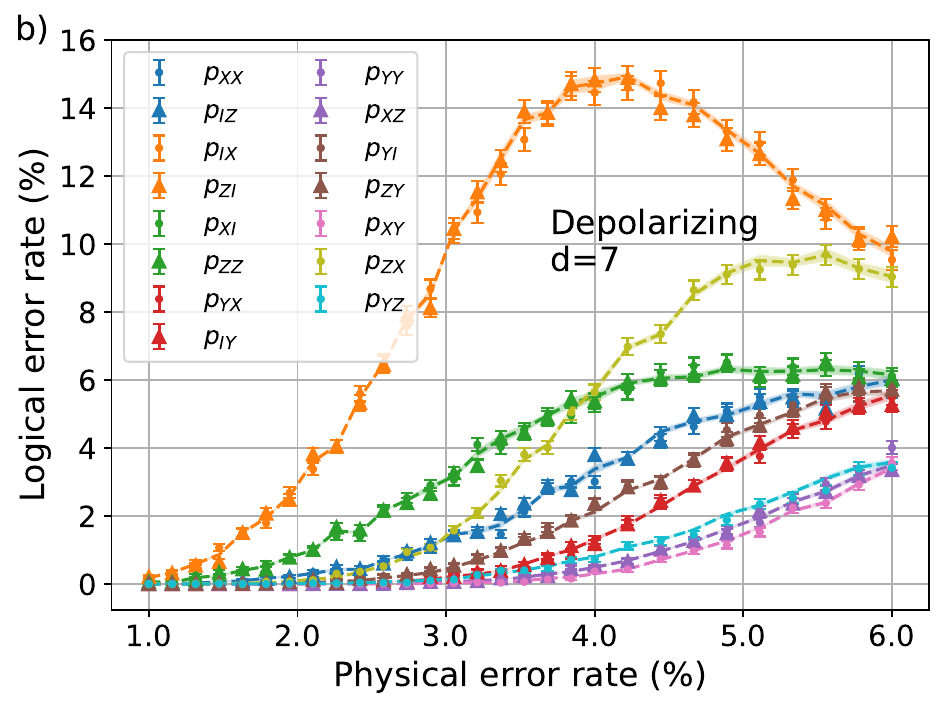}
    \caption{Logical Pauli error rates in CNOT as a function of the physical noise parameter for patch size $d=7$, with (a) independent Pauli $X$ and $Z$ errors with equal error rates, and (b) depolarizing errors on the physical qubits. Numerically obtained error rates are shown with symbols with errorbars. Rates of symmetry partners have the same color; they are in most cases equal to good precision. Expectations for these rates, calculated from the fit parameters $p_1, p_2, p_3$, as described in the main text, are shown with dashed lines, with shading indicating expected uncertainty.}
    \label{fig:table_values_d7}
\end{figure}

With extensive numerical work, we fully characterized the logical error channel of the CNOT for a wide range of code distances and physical error rates. We constructed the $X$- and $Z$-check graphs of the logical CNOT gate, and solved the decoding problem with minimum weight perfect matching decoder (PyMatching) \cite{higgott2022pymatching,higgott2023sparse} under independent and depolarizing noise. To obtain logical error probabilities $p_{IX}, \ldots, p_{XZ}, \ldots$, we ran Monte Carlo simulations, where in each round we identified a logical Pauli error based on the parities of error + correction string endings at boundaries using Table.~\ref{tab:Pauli_strings_classes}.

We chose $h_2=d$ and $h_1=w=1$ to ensure the fault distance to be $d$ with minimal resources. This choice might not the optimal for finite physical error rates as we showed in the previous section.

We found in all cases with independent noise that the logical error channel has the symmetry property expected based on the spacetime diagram, as per Fig.~\ref{fig:table_heatmap}, to a good precision. A numerical example for patch size $d=5$ is also shown in Fig.~\ref{fig:table_heatmap}. For a more systematic exploration, we used a two-step procedure: 
1) we first extracted the string connection probability parameters $p_1$, $p_2$, $p_3$ for each physical error probability 
by fitting a function in the form of Fig.~\ref{fig:table_heatmap}: two-qubit Pauli errors as arguments, and $p_1$, $p_2$ and $p_3$ as fitting parameters (obtaining the logical channel with the structure of Fig.~\ref{fig:table_heatmap} closest to the numerical values); 2) we then compared the numerically obtained values of the frequency of two-qubit errors with the values obtained from the fitted model. 
We show this comparison for patch size $d=7$ in Fig.~\ref{fig:table_values_d7}, and for other parameter sets in Appendix~\ref{sec:CNOT_all_result}.

We found that also with depolarizing errors, the logical error channel of the CNOT can be described with $p_1,p_2,p_3$ in the same way as per Fig.\ref{fig:table_heatmap}. We did not expect this, since the connection of Sec.~\ref{subsec:Symmetry} between the error string probabilities and Pauli errors holds for independent $X$ and $Z$ errors with equal rates, while for depolarizing noise these error are not independent. One possible reason for this factorized structure could be that the independent decoding of $X$ and $Z$ check graphs makes the logical error channel effectively more and more independent as the bulk size grows. We discuss this in more detail in Appendix~\ref{sec:correlation_mwpm}. 

\subsection{Threshold behaviour}

\begin{figure}[!h]
    \centering
    \includegraphics[width = .9\columnwidth]{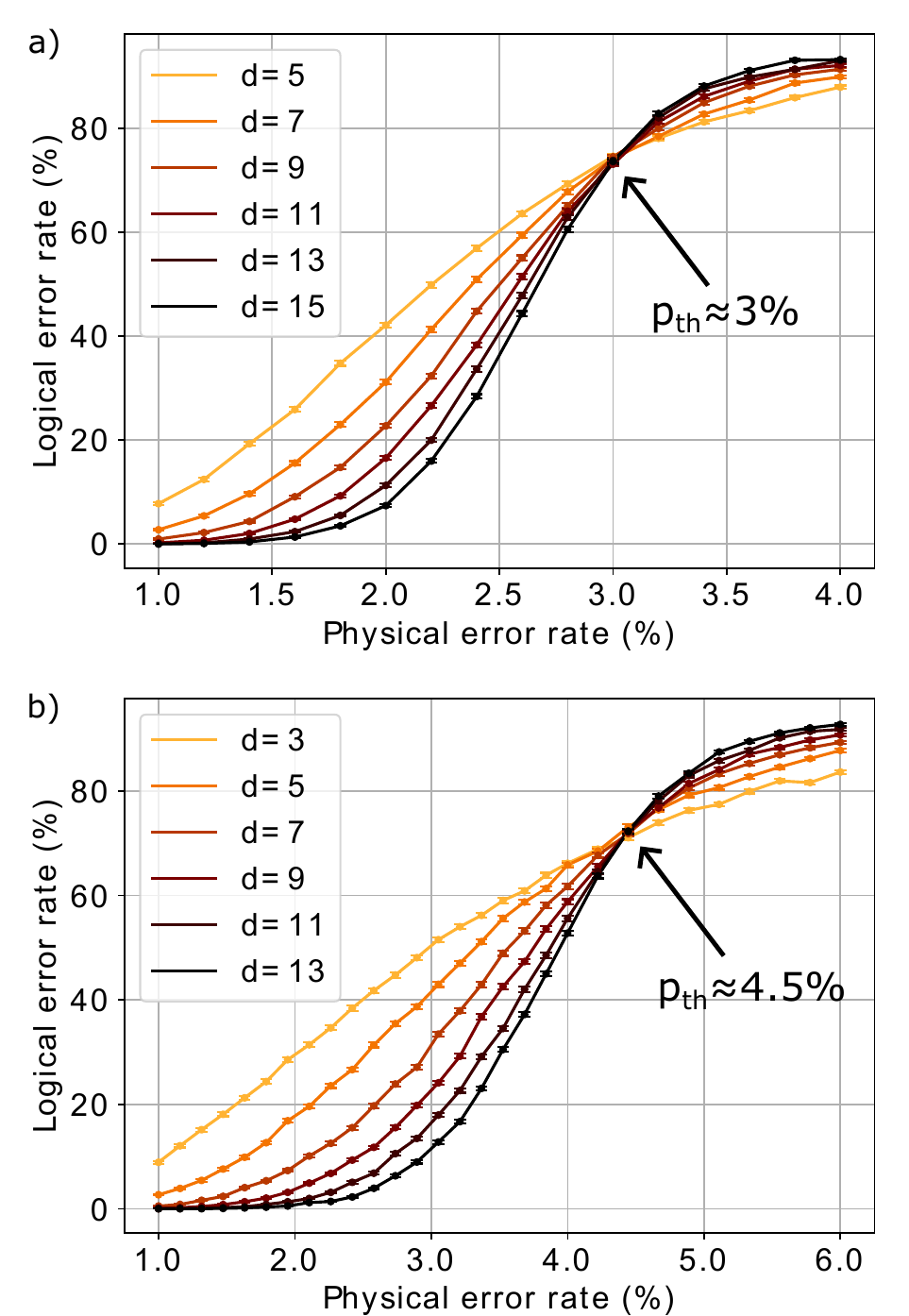}
    \caption{The logical error rates as the function of physical error rates for different fault distances. (a) Independent noise described in Eq.~\eqref{eq:indep_channel}. (b) Depolarizing noise described in Eq,~\eqref{eq:depol_channel}. In both cases we see that the lattice-surgery-based CNOT protocol shows threshold behaviour. The parameters of the simulation are $d=h_2$, $h_1=1$, $w=1$.}
    \label{fig:threshold_indep}
\end{figure}

The surface code, when used as a quantum memory, has a threshold behaviour. There is a critical physical error rate, e.g., $p_{\text{th}}\approx 3\%$ for independent noise \cite{Wang_2003}, below which the
logical error rates can be suppressed exponentially by increasing the code
distance. 

Any lattice-surgery-based logical quantum circuit should also have an error threshold, with the same value as the threshold defined above. Thus, when scaling up the size (fault distance) of a logical circuit, the
logical error rates should decrease whenever the physical error rates are below threshold:    
in the $d\longrightarrow\infty$ limit the classical statistical mechanical model describing the decoding procedure \cite{Wang_2003} is independent of the details of the boundaries as is dominated by the bulk.

We calculated numerically the threshold of the lattice-surgery-based CNOT gate, and found good agreement with the known threshold of the surface code, as shown in  
Fig.~\ref{fig:threshold_indep}.

Although the threshold value is promising for fault-tolerant quantum computation, the error rates here are much higher  than in a simple memory experiment. 
This is due to the higher number of possible error classes, as has been observed for several other logical operations in \cite{Bombin_2023}. 

The threshold value we obtain for depolarizing noise is around $4.5\%$, which is $1.5 $ times the threshold for independent noise. We believe that this is a consequence of the independent decoding of $X$ and $Z$ syndrome graphs. We discuss this in more detail in Appendix.\ref{sec:correlation_mwpm}.

\section{Conclusion}

In this work, we investigated the performance of the lattice-surgery-based logical CNOT protocol between two surface code patches under phenomenological noise. We optimized the number of stabilizer measurement rounds during a lattice surgery protocol realizing the measurement of a two-qubit logical Pauli operator. We found that for equally strong timelike and spacelike errors the optimal number can slightly differ from the naively expected $d$ (code distance) rounds. This is because for small but finite error rates the $1/\text{log}$ correction - coming from the entropy difference of logical error classes - cannot be neglected. We also found that for biased timelike and spacelike errors, the optimal number of stabilizer measurement rounds can heavily differ from $d$. Our findings suggest that it is beneficial to tailor the number of stabilizer measurement rounds during a lattice surgery protocol to the noise model.

We fully characterized the two-qubit Pauli noise channel of the lattice-surgery-based CNOT protocol and found that the symmetry of this protocol appears in the noise channel. We identified symmetry partners among two-qubit logical Pauli errors, which have to have the same probability in the case of equally strong $X$ and $Z$ errors on the physical level. Moreover, the logical error channel can be described with just three logical error parameters. This last result is exact for independent ($X$ and $Z$ errors are uncorrelated) noise, and a pretty good approximation for depolarizing noise. 

A natural next step of our work would be circuit-level simulation of the lattice-surgery-based CNOT protocol with a specific syndrome extraction circuit. With error models close to experimental reality the symmetry of the spacetime diagram will be broken, but only at the boundaries, thus we expect that the logical error channel will be approximately symmetric. Bias in the physical noise (e.g., larger $Z$ than $X$ error rates) will definitely break the symmetry; it is an open question whether symmetrized (e.g., XZZX \cite{Bonilla_Ataides_2021}) surface code restores the symmetry structure. The more general relations between symmetries of logical protocols and the structure of the logical noise also remained unexplored.

Our study can help to understand the structure of the logical noise in surface code-based fault-tolerant quantum computation. Thus, it may have important implications, which have to be considered during the designing of fault-tolerant quantum algorithms. 

\section{Acknowledgement}
This research was supported by the Ministry of Culture and Innovation and the National Research, Development and Innovation Office within the Quantum Information National Laboratory of Hungary (Grant No. 2022-2.1.1-NL-2022-00004). Supported by the Horizon Europe research and innovation programme of the European Union through the HORIZON-CL4-2022- QUANTUM01-SGA project 101113946 OpenSuperQPlus100 of the EU Flagship on Quantum Technologies. We acknowledge the support of the Wigner Research Centre for Physics’ trainee program.

\bibliography{main}

\begin{thebibliography}{10}

\bibitem{kitaev2003fault}
A.Yu. Kitaev.
\newblock ``Fault-tolerant quantum computation by anyons''.
\newblock \href{https://dx.doi.org/10.1016/s0003-4916(02)00018-0}{Annals of Physics {\bf 303}, 2–30}~(2003).

\bibitem{fowler2012surface}
Austin~G. Fowler, Matteo Mariantoni, John~M. Martinis, and Andrew~N. Cleland.
\newblock ``Surface codes: Towards practical large-scale quantum computation''.
\newblock \href{https://dx.doi.org/10.1103/physreva.86.032324}{Physical Review A{\bf 86}}~(2012).

\bibitem{google2023suppressing}
Rajeev~Acharya et~al.
\newblock ``Suppressing quantum errors by scaling a surface code logical qubit''.
\newblock \href{https://dx.doi.org/10.1038/s41586-022-05434-1}{Nature {\bf 614}, 676 -- 681}~(2022).

\bibitem{Bluvstein_2022}
Dolev Bluvstein, Harry Levine, Giulia Semeghini, Tout~T. Wang, Sepehr Ebadi, Marcin Kalinowski, Alexander Keesling, Nishad Maskara, Hannes Pichler, Markus Greiner, Vladan Vuleti{\'{c}}, and Mikhail~D. Lukin.
\newblock ``A quantum processor based on coherent transport of entangled atom arrays''.
\newblock \href{https://dx.doi.org/10.1038/s41586-022-04592-6}{Nature {\bf 604}, 451--456}~(2022).

\bibitem{Marques_2021}
J.~F. Marques, B.~M. Varbanov, M.~S. Moreira, H.~Ali, N.~Muthusubramanian, C.~Zachariadis, F.~Battistel, M.~Beekman, N.~Haider, W.~Vlothuizen, A.~Bruno, B.~M. Terhal, and L.~DiCarlo.
\newblock ``Logical-qubit operations in an error-detecting surface code''.
\newblock \href{https://dx.doi.org/10.1038/s41567-021-01423-9}{Nature Physics {\bf 18}, 80--86}~(2021).

\bibitem{krinner2022realizing}
Sebastian Krinner, Nathan Lacroix, Ants Remm, Agustin Di~Paolo, Elie Genois, Catherine Leroux, Christoph Hellings, Stefania Lazar, Francois Swiadek, Johannes Herrmann, et~al.
\newblock ``Realizing repeated quantum error correction in a distance-three surface code''.
\newblock \href{https://dx.doi.org/10.1038/s41586-022-04566-8}{Nature {\bf 605}, 669--674}~(2022).

\bibitem{Zhao_2022}
Youwei Zhao et~al.
\newblock ``Realization of an error-correcting surface code with superconducting qubits''.
\newblock \href{https://dx.doi.org/10.1103/physrevlett.129.030501}{Physical Review Letters{\bf 129}}~(2022).

\bibitem{Bluvstein_2023}
Dolev Bluvstein et~al.
\newblock ``Logical quantum processor based on reconfigurable atom arrays''.
\newblock \href{https://dx.doi.org/10.1038/s41586-023-06927-3}{Nature}~(2023).

\bibitem{zhang2024demonstrating}
Jiaxuan Zhang, Zhao-Yun Chen, Yun-Jie Wang, Bin-Han Lu, Hai-Feng Zhang, Jia-Ning Li, Peng Duan, Yu-Chun Wu, and Guo-Ping Guo.
\newblock ``Demonstrating a universal logical gate set on a superconducting quantum processor''~(2024).
\newblock  \href{http://arxiv.org/abs/2405.09035}{arXiv:2405.09035}.

\bibitem{hetenyi2024creating}
Bence Het{\'e}nyi and James~R Wootton.
\newblock ``Creating entangled logical qubits in the heavy-hex lattice with topological codes''~(2024).
\newblock  \href{http://arxiv.org/abs/2404.15989}{arXiv:2404.15989}.

\bibitem{menendez2023implementing}
Daniel~Honciuc Menendez, Annie Ray, and Michael Vasmer.
\newblock ``Implementing fault-tolerant non-clifford gates using the [[8, 3, 2]] color code''~(2023).
\newblock  \href{http://arxiv.org/abs/2309.08663}{arXiv:2309.08663}.

\bibitem{Brown_2017}
Benjamin~J. Brown, Katharina Laubscher, Markus~S. Kesselring, and James~R. Wootton.
\newblock ``Poking holes and cutting corners to achieve {C}lifford gates with the surface code''.
\newblock \href{https://dx.doi.org/10.1103/physrevx.7.021029}{Physical Review X{\bf 7}}~(2017).

\bibitem{Shor_1996}
P.W. Shor.
\newblock ``Fault-tolerant quantum computation''.
\newblock In Proceedings of 37th Conference on Foundations of Computer Science.
\newblock \href{https://dx.doi.org/10.1109/SFCS.1996.548464}{Pages 56--65}.
\newblock ~(1996).

\bibitem{beenakker2004charge}
CWJ Beenakker, DP~DiVincenzo, C~Emary, and M~Kindermann.
\newblock ``Charge detection enables free-electron quantum computation''.
\newblock \href{https://dx.doi.org/10.1103/physrevlett.93.020501}{Physical Review Letters {\bf 93}, 020501}~(2004).

\bibitem{Horsman_2012}
Dominic Horsman, Austin~G Fowler, Simon Devitt, and Rodney~Van Meter.
\newblock ``Surface code quantum computing by lattice surgery''.
\newblock \href{https://dx.doi.org/10.1088/1367-2630/14/12/123011}{New Journal of Physics {\bf 14}, 123011}~(2012).

\bibitem{Litinski_2019}
Daniel Litinski.
\newblock ``A game of surface codes: Large-scale quantum computing with lattice surgery''.
\newblock \href{https://dx.doi.org/10.22331/q-2019-03-05-128}{Quantum {\bf 3}, 128}~(2019).

\bibitem{Bravyi_2005}
Sergey Bravyi and Alexei Kitaev.
\newblock ``Universal quantum computation with ideal {C}lifford gates and noisy ancillas''.
\newblock \href{https://dx.doi.org/10.1103/physreva.71.022316}{Physical Review A{\bf 71}}~(2005).

\bibitem{Gupta_2024}
Riddhi~S. Gupta, Neereja Sundaresan, Thomas Alexander, Christopher~J. Wood, Seth~T. Merkel, Michael~B. Healy, Marius Hillenbrand, Tomas Jochym-O’Connor, James~R. Wootton, Theodore~J. Yoder, Andrew~W. Cross, Maika Takita, and Benjamin~J. Brown.
\newblock ``Encoding a magic state with beyond break-even fidelity''.
\newblock \href{https://dx.doi.org/10.1038/s41586-023-06846-3}{Nature {\bf 625}, 259–263}~(2024).

\bibitem{Campbell_2017}
Earl~T. Campbell, Barbara~M. Terhal, and Christophe Vuillot.
\newblock ``Roads towards fault-tolerant universal quantum computation''.
\newblock \href{https://dx.doi.org/10.1038/nature23460}{Nature {\bf 549}, 172--179}~(2017).

\bibitem{Brown_2020}
Benjamin~J. Brown.
\newblock ``A fault-tolerant non-{C}lifford gate for the surface code in two dimensions''.
\newblock \href{https://dx.doi.org/10.1126/sciadv.aay4929}{Science Advances {\bf 6}, eaay4929}~(2020).

\bibitem{Laubscher_2019}
Katharina Laubscher, Daniel Loss, and James~R. Wootton.
\newblock ``Universal quantum computation in the surface code using non-{A}belian islands''.
\newblock \href{https://dx.doi.org/10.1103/physreva.100.012338}{Physical Review A{\bf 100}}~(2019).

\bibitem{bombin2015gauge}
H{\'e}ctor Bomb{\'\i}n.
\newblock ``Gauge color codes: optimal transversal gates and gauge fixing in topological stabilizer codes''.
\newblock \href{https://dx.doi.org/10.1088/1367-2630/17/8/083002}{New Journal of Physics {\bf 17}, 083002}~(2015).

\bibitem{choi2023fault}
Hyeongrak Choi, Frederic~T. Chong, Dirk Englund, and Yongshan Ding.
\newblock ``Fault tolerant non-{C}lifford state preparation for arbitrary rotations''~(2023).
\newblock  \href{http://arxiv.org/abs/2303.17380}{arXiv:2303.17380}.

\bibitem{Beverland_2021}
Michael~E. Beverland, Aleksander Kubica, and Krysta~M. Svore.
\newblock ``Cost of universality: A comparative study of the overhead of state distillation and code switching with color codes''.
\newblock \href{https://dx.doi.org/10.1103/prxquantum.2.020341}{{PRX} Quantum{\bf 2}}~(2021).

\bibitem{Litinski_2018}
Daniel Litinski and Felix~von Oppen.
\newblock ``Lattice surgery with a twist: Simplifying {C}lifford gates of surface codes''.
\newblock \href{https://dx.doi.org/10.22331/q-2018-05-04-62}{Quantum {\bf 2}, 62}~(2018).

\bibitem{Bombin_2007}
H.~Bombin and M.~A. Martin-Delgado.
\newblock ``Optimal resources for topological two-dimensional stabilizer codes: Comparative study''.
\newblock \href{https://dx.doi.org/10.1103/physreva.76.012305}{Physical Review A{\bf 76}}~(2007).

\bibitem{McEwen_2023}
Matt McEwen, Dave Bacon, and Craig Gidney.
\newblock ``Relaxing hardware requirements for surface code circuits using time-dynamics''.
\newblock \href{https://dx.doi.org/10.22331/q-2023-11-07-1172}{Quantum {\bf 7}, 1172}~(2023).

\bibitem{Gidney_2021_stim}
Craig Gidney.
\newblock ``Stim: a fast stabilizer circuit simulator''.
\newblock \href{https://dx.doi.org/10.22331/q-2021-07-06-497}{Quantum {\bf 5}, 497}~(2021).

\bibitem{dennis2002topological}
Eric Dennis, Alexei Kitaev, Andrew Landahl, and John Preskill.
\newblock ``Topological quantum memory''.
\newblock \href{https://dx.doi.org/10.1063/1.1499754}{Journal of Mathematical Physics {\bf 43}, 4452–4505}~(2002).

\bibitem{fowler2014minimum}
Austin~G. Fowler.
\newblock ``Minimum weight perfect matching of fault-tolerant topological quantum error correction in average $o(1)$ parallel time''~(2014).
\newblock  \href{http://arxiv.org/abs/1307.1740}{arXiv:1307.1740}.

\bibitem{edmonds_1965}
Jack Edmonds.
\newblock ``Paths, trees, and flowers''.
\newblock \href{https://dx.doi.org/10.4153/CJM-1965-045-4}{Canadian Journal of Mathematics {\bf 17}, 449–467}~(1965).

\bibitem{higgott2022pymatching}
Oscar Higgott.
\newblock ``Pymatching: A python package for decoding quantum codes with minimum-weight perfect matching''.
\newblock \href{https://dx.doi.org/10.1145/3505637}{ACM Transactions on Quantum Computing {\bf 3}, 1--16}~(2022).

\bibitem{higgott2023sparse}
Oscar Higgott and Craig Gidney.
\newblock ``Sparse blossom: correcting a million errors per core second with minimum-weight matching''~(2023).
\newblock  \href{http://arxiv.org/abs/2303.15933}{arXiv:2303.15933}.

\bibitem{data}
``Numerical data and {P}ython codes''.
\newblock \url{https://doi.org/10.5281/zenodo.11079788}.

\bibitem{Bombin_2023}
Héctor Bombín, Chris Dawson, Ryan~V. Mishmash, Naomi Nickerson, Fernando Pastawski, and Sam Roberts.
\newblock ``Logical blocks for fault-tolerant topological quantum computation''.
\newblock \href{https://dx.doi.org/10.1103/prxquantum.4.020303}{PRX Quantum{\bf 4}}~(2023).

\bibitem{gidney2023inplace}
Craig Gidney.
\newblock ``Inplace access to the surface code y basis''.
\newblock \href{https://dx.doi.org/10.22331/q-2024-04-08-1310}{Quantum {\bf 8}, 1310}~(2024).

\bibitem{gehér2023errorcorrected}
György~P. Gehér, Campbell McLauchlan, Earl~T. Campbell, Alexandra~E. Moylett, and Ophelia Crawford.
\newblock ``Error-corrected hadamard gate simulated at the circuit level''.
\newblock \href{https://dx.doi.org/10.22331/q-2024-07-02-1394}{Quantum {\bf 8}, 1394}~(2024).

\bibitem{Chamberland_2022_1}
Christopher Chamberland and Earl~T. Campbell.
\newblock ``Universal quantum computing with twist-free and temporally encoded lattice surgery''.
\newblock \href{https://dx.doi.org/10.1103/prxquantum.3.010331}{{PRX} Quantum{\bf 3}}~(2022).

\bibitem{Chamberland_2022_2}
Christopher Chamberland and Earl~T. Campbell.
\newblock ``Circuit-level protocol and analysis for twist-based lattice surgery''.
\newblock \href{https://dx.doi.org/10.1103/physrevresearch.4.023090}{Physical Review Research{\bf 4}}~(2022).

\bibitem{Bonilla_Ataides_2021}
J.~Pablo Bonilla~Ataides, David~K. Tuckett, Stephen~D. Bartlett, Steven~T. Flammia, and Benjamin~J. Brown.
\newblock ``The {XZZX} surface code''.
\newblock \href{https://dx.doi.org/10.1038/s41467-021-22274-1}{Nature Communications{\bf 12}}~(2021).

\bibitem{Litinski_2017}
Daniel Litinski and Felix von Oppen.
\newblock ``Braiding by {M}ajorana tracking and long-range {CNOT} gates with color codes''.
\newblock \href{https://dx.doi.org/10.1103/physrevb.96.205413}{Physical Review B{\bf 96}}~(2017).

\bibitem{Wang_2003}
Chenyang Wang, Jim Harrington, and John Preskill.
\newblock ``Confinement-{H}iggs transition in a disordered gauge theory and the accuracy threshold for quantum memory''.
\newblock \href{https://dx.doi.org/10.1016/s0003-4916(02)00019-2}{Annals of Physics {\bf 303}, 31--58}~(2003).

\end{thebibliography}
\bibliographystyle{quantum}

\appendix


\section{Numerical results for the logical CNOT gate}
\label{sec:CNOT_all_result}
To investigate the $\Bar{Z}\Bar{Z}$-measurement and the logical CNOT operation, we have performed Pauli frame simulations. We have constructed check graphs associated with the spacetime diagrams of the experiments testing the $\Bar{Z}\Bar{Z}$-measurement and the CNOT gate which we could put into Pymatching  \cite{higgott2022pymatching, higgott2023sparse}. With the correction operator and the randomly generated error history in hand, we could identify the specific logical error mechanism that occurred during the logical operation by checking the parities of error + correction string endings at boundaries. To obtain a statistically significant data on these logical error mechanisms multiple Monte Carlo rounds are needed. For the CNOT gate, we collected data from $10^4$ Monte Carlo rounds. In case of the $\Bar{Z}\Bar{Z}$ measurement, we investigated lower error rates, therefore $10^4$ rounds would not have been enough. So we simulated until we reached 10 erroneous logical operations for which depending on the size of the check graph $10^5-10^7$ rounds were enough. 

\begin{figure*}[!ht]
    \centering
    \includegraphics[width=0.48\textwidth]{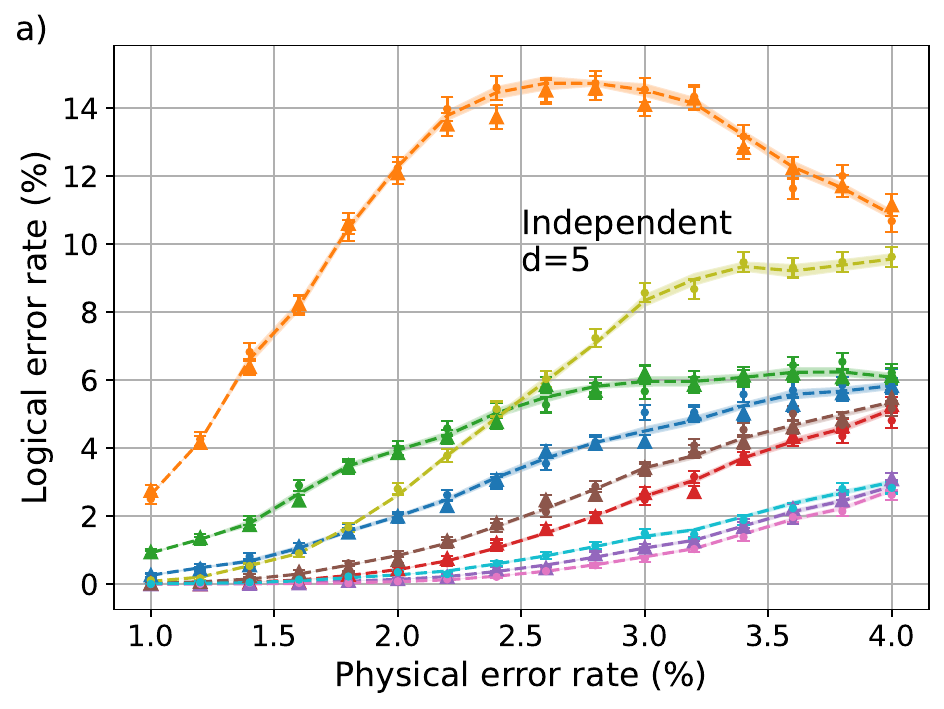}
    \includegraphics[width=0.48\textwidth]{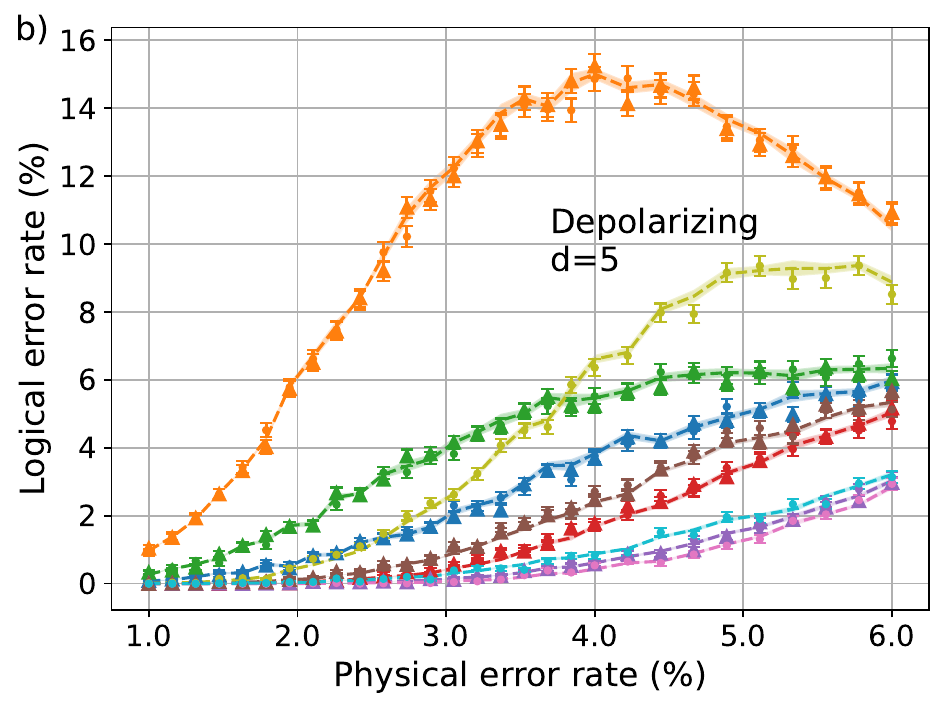}
    \includegraphics[width=0.48\textwidth]{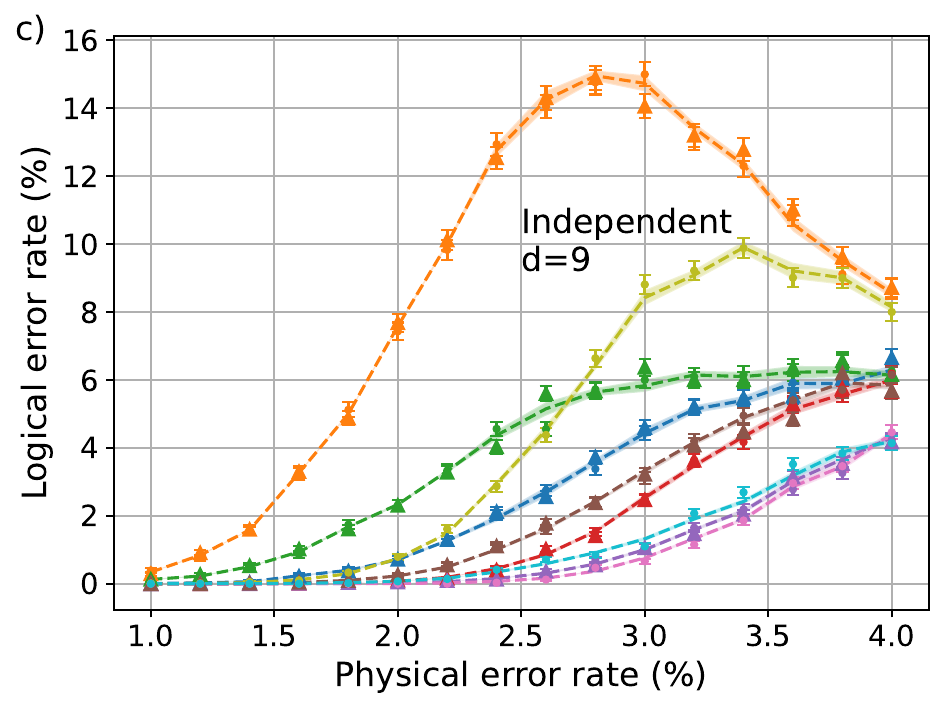}
    \includegraphics[width=0.48\textwidth]{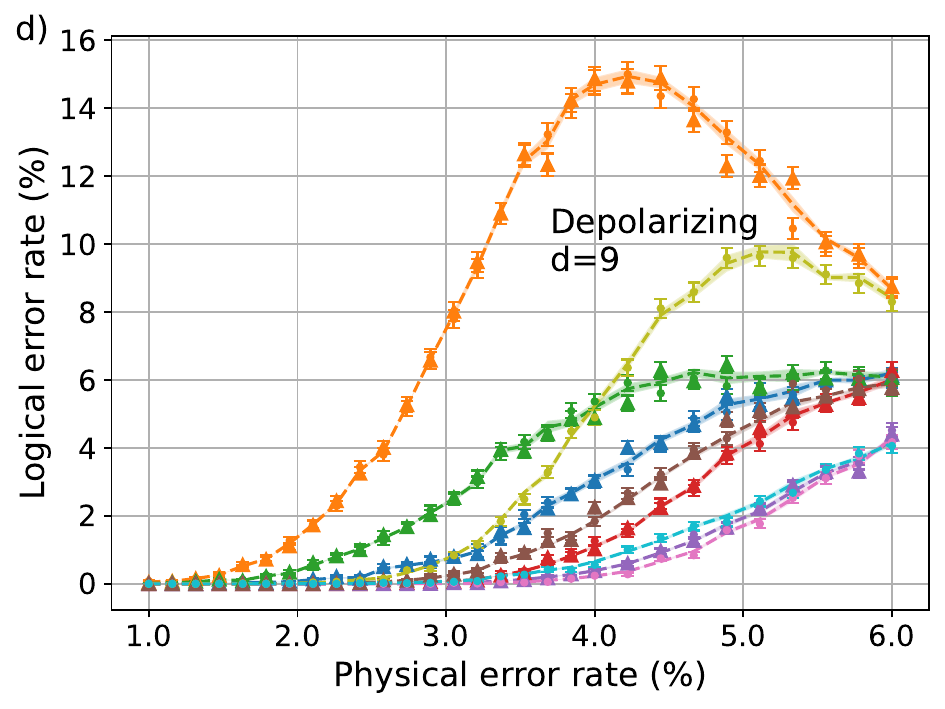}
    \includegraphics[width=0.48\textwidth]{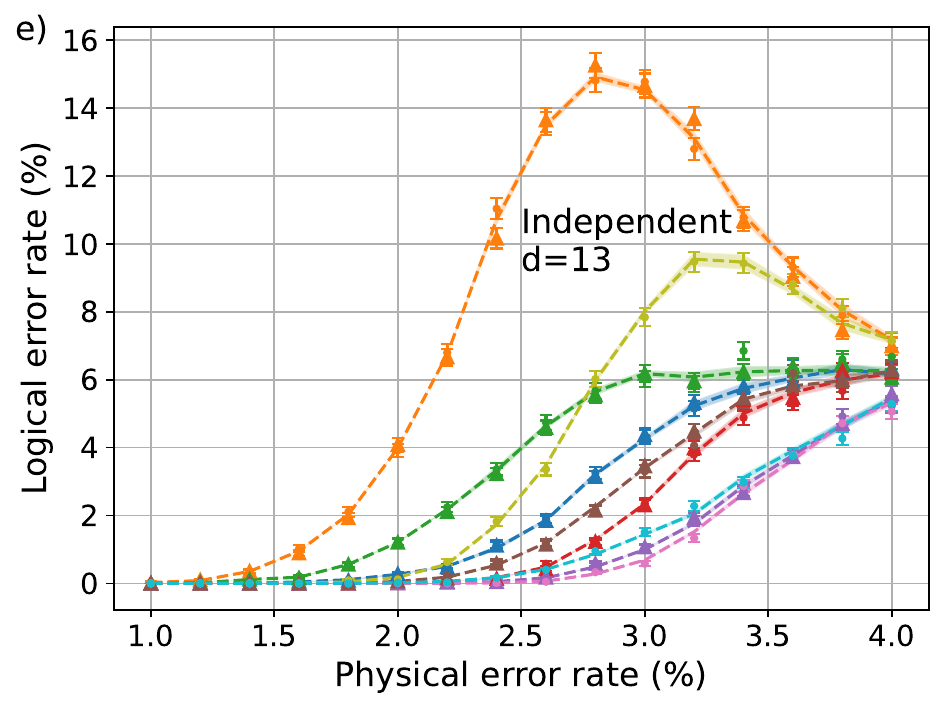}
    \includegraphics[width=0.48\textwidth]{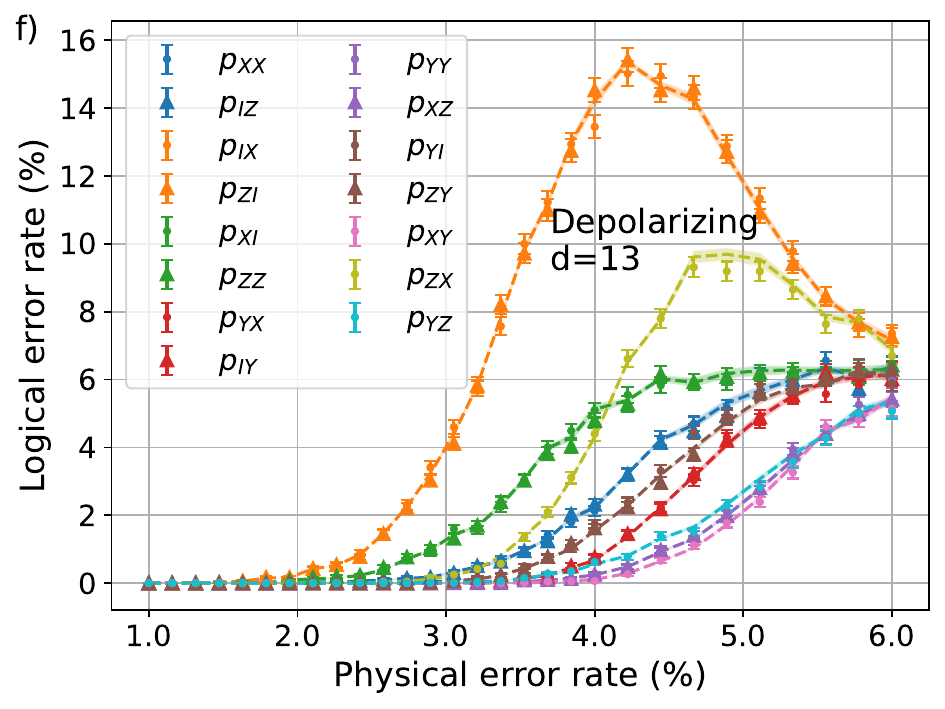} 
    \caption{Logical Pauli error rates in CNOT as a function of the physical noise parameter for various patch sizes $d$, with independent Pauli $X$ and $Z$ errors with equal error rates, or depolarizing errors on the physical qubits, as indicated on the plots. Numerically obtained error rates are shown with symbols with errorbars. Rates of symmetry partners have the same color; they are in most cases equal to good precision. Expectations for these rates, calculated from the fit parameters $p_1, p_2, p_3$, as described in the main text, are shown with dashed lines, with shading indicating expected uncertainty. The other geometrical parameters of the procedure for all patch sizes are: $w=1$, $h_1=h_3=1$, $h_2=d$.}
    %
    \label{fig:CNOT_error_connections}
\end{figure*}

We collected Monte Carlo data for CNOT with geometrical parameters $d=5, 7, 9, 11, 13$, $w=1$, $h_1=h_3=1$, $h_2=d$ under independent noise or depolarizing noise with various error rates. For each simulation setting we determined the $4\times4$ table representing the two qubit logical Pauli channel. We depict the numerically obtained values in Fig.~\ref{fig:CNOT_error_connections} in the same way as we did for $d=7$ in Fig.~\ref{fig:table_values_d7} for all other patch sizes. The data is available at \cite{data}.
Our findings for $d=7$ also hold for other patch sizes.

In Fig.~\ref{fig:CNOT_error_connections} one may notice that the $IX$ and $ZI$ outcomes have much higher probabilities than the others. 
This is because the corresponding logical error strings connect the two largest $X$ or $Z$ boundaries (while there is no connection between separated boundaries on the other check graph). Therefore these classes of connections have high multiplicities, which leads to them dominating the logical errors for small error rates.

\section{Correlations in the logical error channel under minimum weight perfect matching decoding}
\label{sec:correlation_mwpm}

In Sec.~\ref{sec:CNOT_characterization} we saw that the two-qubit logical Pauli channel for independent noise with equal $X$ and $Z$ error rates can be fully described by the three string connection probability parameters $p_1$, $p_2$, $p_3$. The numerics reported in Sec.\ref{sec:CNOT_characterization} has also shown that this decomposition also works for the depolarizing noise channel -- surprising, since here the $X$ and $Z$ errors are correlated. We think that this is due to the use of the minimum weight perfect matching decoding, under which the correlations between $X$ and $Z$ errors are suppressed on the logical level.

The minimum weight perfect matching decoder finds the most probable $X$-type and $Z$-type error histories independently, that are consistent with the given syndrome. 
This error history will not be the most probable error history necessarily if correlations between $X$ and $Z$ errors are present. Therefore, we assume that for low enough error rates these correlation will not affect the final logical error channel.  

We find that the correlations in the logical noise channel are disappearing as we are scaling up the system size. To characterize the strength of correlations in the logical error channel of single logical qubit we introduce the following measure:
\begin{equation} \label{eq:measure}
    M = \frac{|P_IP_Y - P_XP_Z|}{P_X + P_Y + P_Z}\ ,
\end{equation}
where the logical error channel is defined as:
\begin{equation}
    \varepsilon(\rho)=P_I\rho+P_Z\Bar{Z}\rho\Bar{Z}+P_Y\Bar{Y}\rho\Bar{Y}+P_X\Bar{X}\rho\Bar{X}.
\end{equation}

\begin{figure*}[!ht]
    \centering
    \begin{subfigure}[b]{0.48\textwidth}
    \centering
    \includegraphics[width=\textwidth]{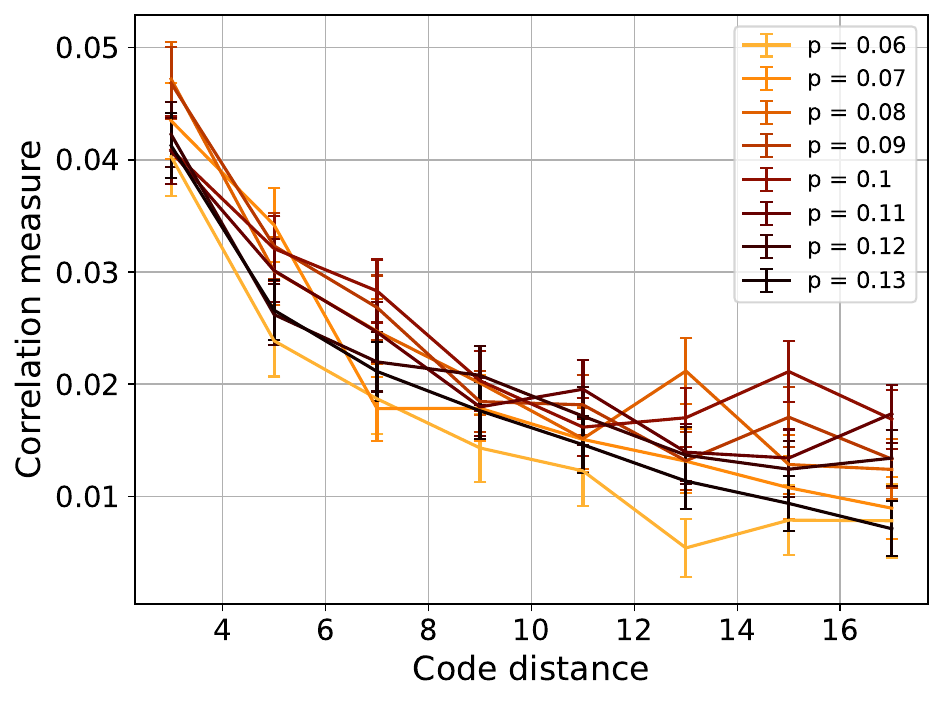}
    \caption{}
    \end{subfigure}
    \quad
    \begin{subfigure}[b]{0.48\textwidth}
    \centering
    \includegraphics[width=\textwidth]{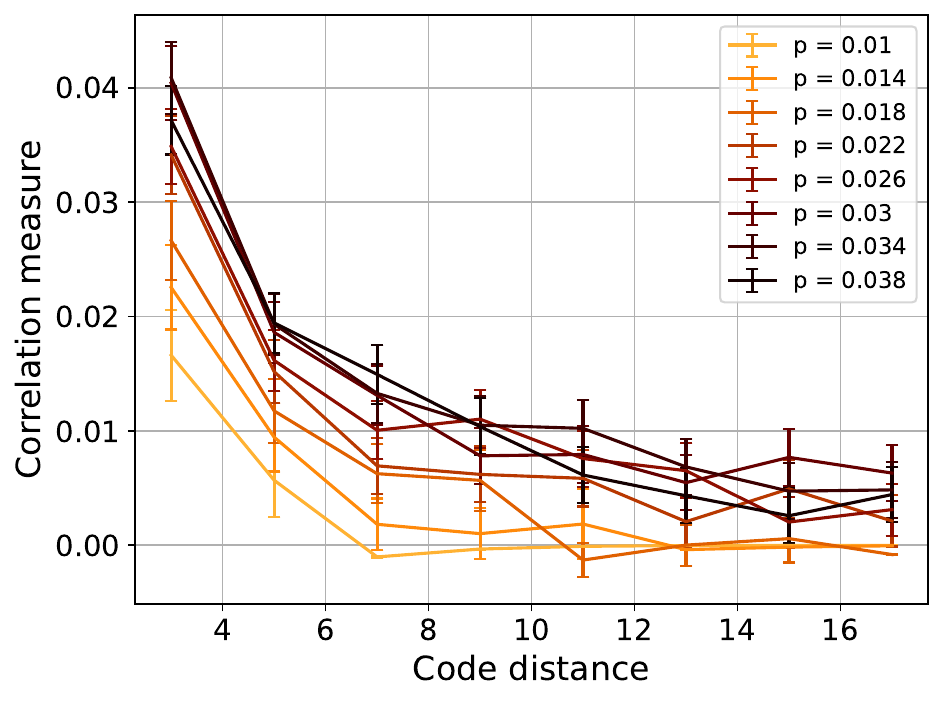}
    \caption{}
    \end{subfigure}
    \caption{Correlations between the possible Pauli outcomes of a memory experiment under depolarizing noise with minimum weight decoder. a) Surface code with error correction after one round of stabilizer measurements ($d \times d \times 1$). b) Surface code with error correction after $d$ round of stabilizer measurements ($d\times d\times d$). As the bulk size grows the logical error channel is becoming more and more similar to independent noise due to the independence in the decoding protocol. }
    \label{fig:memory_experiments}
    
\end{figure*}



\end{document}